\documentclass[12pt]{article}
\pdfoutput=1
\usepackage{amsmath, amssymb}    
\usepackage{graphicx}
\usepackage{dcolumn}
\usepackage{bm}
\usepackage{hyperref}
\usepackage{color}

\newcommand{\be}{\begin{equation} } 
\newcommand{\ee}{\end{equation} } 
\newcommand{\ba}{\begin{array} } 
\newcommand{\ea}{\end{array} } 
\newcommand{\bear}{\begin{eqnarray} } 
\newcommand{\eear}{\end{eqnarray} }

\newcommand{\thc}{\Theta_{_{\! \rm C}}}  
\newcommand{\thi}{\Theta_{_{\! \rm I}}}
\newcommand{\thr}{\Theta_{_{\! \rm R}}}

\newcommand{\tht}{\Theta_{_{\! \rm T}}}

\title{
\vspace*{-3.4cm}
\begin{flushright}
\normalsize{ \small   Fermilab-PUB-25-0816-T }
\end{flushright}
\vspace*{0.99cm}
\large  
{\bf  \ Octet scalars shaping LHC distributions in 4-jet final states  
\\ [9mm]
}
\vspace*{0.3cm}   
}

\author{{\bf  \normalsize 
Bogdan A. Dobrescu and Max H. Fieg} 
\vspace{5mm}
\\
\normalsize\emph{Particle Theory Department, Fermilab, Batavia, IL 60510, USA    
\footnote{bdob@fnal.gov, mfieg@fnal.gov}
}
}
\date{ \normalsize  November 12, 2025 }

\begin{document}  
\setcounter{page}{0}  
\maketitle  

\begin{abstract}  
We study properties of a hypothetical scalar particle,  $\Theta$, which is a color octet and an electroweak singlet. At hadron colliders, $\Theta$ is pair produced through its QCD coupling to gluons, so that its mass determines the cross section. It decays at tree level into $q\bar q$ through dimension-5 operators, and at one loop into gluons. Thus, the main LHC signature of $\Theta$  is a pair of dijets of equal invariant mass. The CMS search in this channel shows a $3.6\sigma$ excess over the QCD background for a dijet mass $M_{jj} \approx 0.95$ TeV,  which can be due to $\Theta$: its production cross section (65 fb for a real scalar) and the acceptance of the CMS event selection applied to $p p \to \Theta \Theta \to \! (q \bar q)(q \bar q)$  yield a rate consistent with the excess. Furthermore, the shape of the $d\sigma/d M_{jj}$ signal is in agreement with the CMS result. Given the data-driven background fit performed by CMS, we find that a complex scalar (whose production rate is twice as large) fits better the data than a real scalar. Besides the pair of dijets,  testable LHC signals include a trijet-dijet topology, a $t\bar t$ pair plus a dijet resonance, as well as final states involving a Higgs, $W$ or $Z$ boson plus jets.
\end{abstract} 

\vfil

\thispagestyle{empty}  
  
\setcounter{page}{1}  
  
\vspace*{0.31cm}    
  
\newpage  
  
\tableofcontents
  
\vspace*{0.31cm}    
  
\baselineskip18pt   

\bigskip\bigskip\bigskip

\section{Introduction} 
\label{sec:intro}

The LHC experiments searching for new phenomena at high transverse momenta, ATLAS and CMS, have made remarkable progresses within the last few years, making them more likely to discover physics beyond the Standard Model (SM). 
Examples of such progresses include better methods of leveraging jet substructure \cite{CMS:2023tlv,ATLAS:2020gwe}, the extensive use of deep-learning neural networks to identify objects \cite{ATLAS:2025dkv, CMS:2022prd}, substantial improvements of the detectors \cite{ATLAS:2023dns,CMS:2023gfb}, and a widening of the types of signals searched for \cite{searches}.

If new particles carry QCD color and have mass near the TeV scale, then they can be copiously produced at the LHC. Even so, their discovery is not guaranteed, because the backgrounds may be large, and difficult to compute. An intriguing type of hypothetical particle with such properties is a color-octet scalar that is an electroweak singlet. 
This particle, with the same SM charges under the $SU(3)_c\times SU(2)_W\times U(1)_Y$ gauge group as the gluon, namely $(8,1,0)$,   
is predicted in various models for physics beyond the Standard Model (SM). 
A color-octet real scalar, labelled here $\Theta$, is included  in the scalar sector associated with the spontaneous breaking of 
an extended QCD gauge group \cite{Bai:2010dj, Chivukula:2013xka, Bai:2018jsr}, or as a Kaluza-Klein mode of the gluon polarization along two extra dimensions \cite{Burdman:2006gy, Dobrescu:2007xf} (referred to as `spinless adjoint'). 
A color-octet complex scalar (which is just a set of two $\Theta$ real scalars), labelled here $\Theta_{_{\! \rm C}}$,
is part of the ${\cal N} =2$ supersymmetric multiplet of the gluon \cite{Fox:2002bu}-\cite{Martin:2019eus},  and is sometimes called `sgluon'.

In this article we study the properties of color-octet real or complex scalars in simple renormalizable extensions of the SM. The main production of $\Theta$ at hadron colliders is in pairs through its coupling to gluons \cite{Dobrescu:2007yp}, which has a cross section entirely determined by the particle mass, $M_\Theta$. 
If no particles other than $\Theta$ exist beyond the SM, then $\Theta$ decays at one loop into two gluons with a highly suppressed width  \cite{Bai:2010dj} and a branching fraction close to 100\%. 

If, however, new particles couple to $\Theta$,  then other decay modes become the predominant ones even when those particles are substantially heavier than $\Theta$. Couplings of $\Theta$ to quark-antiquark pairs are induced by dimension-5 operators, which can be generated (at tree level or through loops) by heavier fermions and bosons with various SM gauge charges. 
In the case of light quarks, $q$, this leads to a $\Theta \to q \bar q$ decay with the most common final state being, as in the case of $\Theta \to gg$, two hadronic jets of high momenta. 

Thus, LHC signals where a $\Theta$ pair is produced through QCD, followed by each scalar decaying to two jets, represent a natural signal of color-octet scalars \cite{Dobrescu:2007yp, Bai:2010dj,  Gerbush:2007fe, Arnold:2011ra}.
Although there are many similarities between the gluon and quark final states, we will show that the processes $pp \to \Theta\Theta \to (q \bar q)(q \bar q) \to 4j$ and $pp \to \Theta\Theta \to (gg)(gg) \to 4j$ produce substantially different shapes of the dijet invariant mass distributions due to the higher QCD radiation from gluons than from quarks.

Similar phenomena occur in the case of the complex scalar $\Theta_{_{\! \rm C}}$, with the main difference that the cross section for $pp \to \Theta_{_{\! \rm C}}\Theta_{_{\! \rm C}}$ is twice as large as the one for $\Theta$ pair production. Another difference is that $\Theta_{_{\! \rm C}}$ can be charged under a global $U(1)$ symmetry that would keep $\Theta_{_{\! \rm C}}$  stable in the absence of other particles beyond the SM.

A CMS search in the dijet pair channel \cite{CMS:2022usq} has reported a $3.6\sigma$ excess of events compared to the QCD background, at an average dijet mass of 950 GeV.
We will show that $\Theta$ pair production has a cross section at that mass consistent with the size of the CMS excess, and furthermore the shape of the average dijet invariant mass distribution provides a good fit to the data in the case of quark final states. 
Given that the signal has the right size without any tunable parameter, it is imperative to investigate, both theoretically and experimentally, the hypothesis of a $\Theta$ particle of mass near 1 TeV.

The QCD background is estimated by CMS using a data-driven method based on a smooth parametrization fitted to the average dijet invariant mass. When extracting the probability of an excess, CMS fitted that background plus a signal obtained from simulating stop pair production. We will demonstrate that the $pp \to \Theta\Theta \to (q \bar q)(q \bar q) \to 4j$ signal produces a different shape than stop pair production, such that doubling its rate provides a slightly better fit to the CMS data. This implies that $\Theta_{_{\! \rm C}}\Theta_{_{\! \rm C}}$ production is favored compared to the SM prediction with a significance above $3.6\sigma$.
Alternative new-physics explanations for this excess have been proposed in \cite{Dobrescu:2024mdl, Crivellin:2022nms}.

The phenomenology of the color-octet scalars studied here, of mass in the TeV range, would continue to be interesting even if  the current CMS excess were just a fluctuation of the QCD background. 
Studies at the LHC of simple SM extensions with one or two new particles may provide a first glimpse into new physics. Furthermore, the larger data sets expected at the high-luminosity runs of the LHC, combined with much improved detectors and novel experimental methods will allow improved sensitivity to color-octet scalars over a wider range of masses.

In Section \ref{sec:interactions} we discuss the renormalizable Lagrangian that describes $\Theta$, first in the absence of new particles, and then in the presence of some heavier particles that generate the $\Theta$ coupling to quarks. In Section \ref{sec:950GeV} we show that the CMS excess at 950 GeV can be explained by a real scalar $\Theta$. We then consider the case (Section \ref{sec:2ormore}) where the color octet is a complex scalar, which also turns out to provide a better fit to the CMS data.
Our conclusions can be found in Section \ref{sec:conclusions}.

\medskip  

\section{Interactions of a color-octet real scalar}     
\label{sec:interactions}   \setcounter{equation}{0}

Consider a color-octet scalar field $\Theta^a$, where $a = 1, ...,8$ is the index associate with the color degrees of freedom in the adjoint representation of the QCD gauge group $SU(3)_c$. We take $\Theta^a$ to be a real (not complex) scalar field and an electroweak singlet. 

In this Section we analyze the renormalizable Lagrangian that describes the properties of the $\Theta$ particle,
first in the presence of just the SM, and later (Section \ref{sec:SMTc}) including effects of new heavier particles. 

\subsection{$\Theta$ scalar and the SM}
\label{sec:SMT} 

Assuming for now that there are no additional fields beyond the SM, the most general renormalizable Lagrangian terms that contain $\Theta$ are
\be
{\cal L}_\Theta = \frac{1}{2} \left( D^\mu\Theta^a \right) \, \left(D_\mu\Theta^a  \right)  - V(\Theta) - \frac{\lambda_H}{2} H^\dagger H \,  \Theta^a \, \Theta^a ~~,
\label{eq:lagrangian}
\ee
All couplings to gluons are contained in the kinetic term, which is the first term in ${\cal L}_\Theta$, with the covariant derivative given by
\be
D_\mu \Theta^a  = \partial_\mu \Theta^a  + g_s  f_{abc} \, G_\mu^b \, \Theta^c  ~~.
\ee
Here $b,c = 1,...,8$ are color indices, $G_\mu^b$ is the gluon field, and $g_s$ is the QCD gauge coupling. 
The structure constants of $SU(3)$, $f_{abc}$, are totally antisymmetric in the color indices, 
 and are related to the  generators $T^a$ of the fundamental $SU(3)$ representation by 
$f_{abc} = - 2i \, {\rm Tr} \!\left( [ T^a,T^b ] T^c \right)$.

The second term in (\ref{eq:lagrangian}) is the scalar potential for $\Theta$, which includes $SU(3)_c$-invariant mass, trilinear and quartic terms:
\be
V(\Theta) = \frac{1}{2} {\widetilde{M}_\Theta}^{\, 2} \, \Theta^a \, \Theta^a + \mu_\Theta  \, d_{abc} \, \Theta^a  \,  \Theta^b  \,  \Theta^c + 
\frac{\lambda_\Theta}{8} \,  \Theta^a\Theta^a  \, \Theta^b\Theta^b  ~~.
\label{eq:potential}
\ee
The squared-mass and dimensionless quartic coupling satisfy ${\widetilde{M}_\Theta}^{\, 2} > 0$ and $\lambda_\Theta > 0$, which are necessary conditions to avoid a color-breaking VEV for $\Theta$. Through a sign redefinition of $\Theta$, the trilinear coupling (a parameter of mass dimension one) is taken to satisfy $\mu_\Theta > 0$. The $SU(3)$ tensor $d_{abc} = 2 \, {\rm Tr}\! \left( \{ T^a,T^b\} T^c \right)$ is totally symmetric, and its only nonzero elements up to permutations are 
\bear  
 && d_{aa8} = -2 d_{bb8} = - d_{888} = \frac{1}{\sqrt{3}} \; \; , \; \; {\rm for} \; \; a = 1,2,3   \; , \;\; b = 4, 5, 6 , 7 ~~,
\nonumber \\ [-1mm]
 \\  [-1mm]
&& d_{146} = d_{157} = -d_{247} = d_{256} = d_{344} = d_{355} = - d_{366} = - d_{377} = \frac{1}{2}  ~~.
\nonumber
\eear

In the last term of  (\ref{eq:lagrangian}), $H$  is the SM Higgs doublet, and $\lambda_H$ is a real dimensionless coupling.
As $H$ has a VEV fixed by the electroweak scale ($v_H \approx 174$ GeV), it also contributes to the total squared-mass of $\Theta$,
which must be positive so that  $\Theta$ does not acquire a VEV:
\be
M^2_\Theta = {\widetilde{M}_\Theta}^{\, 2} + \frac{\lambda_H}{2}  v_H^2  \, > \, 0 ~~.
\ee
Even when $\lambda_H$ is negative, the above condition is easily satisfied for ${\widetilde{M}_\Theta}^{\, 2} \gg v_H^2$, which is the region of parameter space preferred by phenomenology (see Section \ref{sec:950GeV}). Note that $\Theta$ is a physical field, so its mass satisfies $M_\Theta > 0$.  The last term in (\ref{eq:lagrangian}) implies an interaction of one SM Higgs boson ($h^0$) with two $\Theta$'s, which contributes to $h^0$ production at hadron colliders through gluon fusion \cite{Boughezal:2010ry, Dobrescu:2011aa, Kumar:2012ww}, as well as an $h^0 h^0 \, \Theta \Theta$ interaction that enhances di-Higgs production \cite{Kribs:2012kz}.

There is one last condition that ensures the global minimum of $V(\Theta)$ is at $\langle \Theta^a \rangle = 0$ for each $a=1, ..., 8$, namely that the ratio of the trilinear coupling to $M_\Theta$ is smaller than a certain function of $\lambda_\Theta$ and  $M_\Theta$: 
\be
\frac{\mu_\Theta}{M_\Theta} \leq  f(\lambda_\Theta, M_\Theta)   ~~.
\label{eq:mumax}
\ee
This function increases with both $M_\Theta$ and $\lambda_\Theta$ because a larger positive squared-mass or a larger quartic term disfavors a deeper $V(\Theta)$ minimum away from the origin. 
Our numerical minimization of $V(\Theta)$ gives the following values for this upper limit on $\mu_\Theta/M_\Theta$:
for  $\lambda_\Theta = 0.1$,   $f(0.1, 0.9  \, {\rm TeV}) \approx 0.704$ and $f(0.1, 2  \, {\rm TeV}) \approx 1.43$, 
while for  
$\lambda_\Theta = 1$,   $f(1, 0.9 \,{\rm TeV}) \approx 1.25$ and $f(1, 2  \, {\rm TeV}) \approx 1.74$.
Values of $\lambda_\Theta$ substantially above 1 are problematic as $\lambda_\Theta$ increases with the energy scale such that the theory described by ${\cal L}_\Theta$ becomes ill-defined at a scale not much larger than $M_\Theta$, so that some new physics ({\it e.g.}, $\Theta$ compositeness) would already have important effects.

The only term in (\ref{eq:lagrangian}) that is not invariant under the  $\Theta \to - \Theta$ transformation is the trilinear term from $V(\Theta)$. Consequently, any decay of  $\Theta$ must have a width proportional to an even power of $\mu_\Theta$. By far the dominant decay mode arises at one loop and is into two gluons. Its decay width was computed in \cite{Bai:2010dj},  and is given by
\be
\Gamma (\Theta \to gg) = \frac{15}{2^{11}\pi^5} \left( \frac{\pi^2}{9} - 1 \right)^{\! 2} g_s(M_\Theta)^4 \, \frac{\mu_\Theta^2}{M_\Theta} ~~.
\label{eq:Thetagg}
\ee
The QCD  gauge coupling $g_s$ (related to the strong coupling constant by $g_s = \sqrt{4\pi \alpha_s} \, $) is 
evaluated here at the  $M_\Theta$ scale. Using the SM renormalization group evolution for $g_s$ up to a scale $M_\Theta = 0.9$ TeV gives $g_s(M_\Theta) \approx 1.06$. At larger scales $g_s$ decreases slowly, {\it e.g.}, $M_\Theta = 2$ TeV gives $g_s(M_\Theta) \approx 1.02$.

Besides the loop suppression, there is an unexpectedly small factor (arising from a loop integral explicitly analyzed in \cite{Dobrescu:2011px}) that reduces the width (\ref{eq:Thetagg}): 
\be
\left( \frac{\pi^2}{9} - 1 \right)^{\! 2} \approx 9.34 \times 10^{-3} ~~.
\ee
This extra suppression could be interpreted as a ``geometric fine-tuning", arising from $\pi$ being within $\sim \! 5\%$ of 3. Naive dimensional analysis would not anticipate a width that small. This is a case where mathematics, rather than physics, appears to be finely tuned. The same suppression factor arises in the decay width of a color-octet scalar that is a weak-doublet \cite{Gresham:2007ri} or a weak-triplet \cite{Dobrescu:2011px}; 
incidentally, a $\pi^2/9 -\! 1$ factor (not squared) appears in the width of ortho-positronium \cite{Itzykson:1980rh}. 

The width-to-mass ratio of $\Theta$  is given by 
\be
\frac{1}{M_\Theta} \, \Gamma (\Theta \to gg) \, \approx \, 2.23 \times 10^{-7} \, g(M_\Theta)^4  \left( \frac{\mu_\Theta}{M_\Theta}\right)^{\! 2}  \, < \, 4.7 \times 10^{-7} ~~.
\label{eq:width_gg}
\ee
where the last inequality is obtained for $\lambda_\Theta \leq 1$ in (\ref{eq:mumax}), and for the value of $g_s$ evaluated at $M_\Theta \approx 1$ TeV.
Although the above $\Theta$ width is unexpectedly small, even for a value of the trilinear coupling as small as $\mu_\Theta = 10^{-3} M_\Theta$, the rest-frame decay length of $\Theta$ remains microscopic ($4 \times 10^{-7}$m), so $\Theta$ decays are prompt on current collider detector time scales.

Single-$\Theta$ production at hadron colliders occurs through gluon fusion, due to the same extra-suppressed $\Theta$ loop. Its parton-level production cross section is given at leading order by 
\be
\hat \sigma ( gg \to \Theta) =  \frac{\pi^2}{M_\Theta}  \,  \Gamma (\Theta \to gg)  \;  \delta(\hat s - M^2_\Theta )   ~,
\ee
where $\hat s$ is the center-of-mass energy of the partonic collision. The single-$\Theta$ production  cross section in proton-proton collisions with center-of-mass energy $\sqrt{s}$  is then 
\be
 \sigma ( gg \to \Theta) =  \frac{\pi^2}{M_\Theta \, s}  \,  \Gamma (\Theta \to gg)  \, \int_{M_\Theta^2/s}^1 \frac{dx}{x} \,  g(x)  \; g \!\left(M_\Theta^2/(x s) \right)~~.
\ee
Here $g(x)$ is the parton distribution function (PDF) of the gluon, with $x$ being the ratio of the gluon and proton momenta.
For $M_\Theta = 0.9$ TeV,  $ \sigma ( gg \to \Theta) \approx 6$ fb. For larger  $M_\Theta $ the cross section drops  fast: {\it e.g.}, $M_\Theta = 1.2$ TeV gives  $ \sigma ( gg \to \Theta) \approx 2$ fb. 
These values for the single-$\Theta$ production cross section are three orders of magnitude smaller than the sensitivity of the dijet resonance searches at masses above 0.9 TeV \cite{CMS:2025ffj, ATLAS:2018qto}.
It is noteworthy, though, that interference between the QCD background and the single-$\Theta$ signal results in complicated shapes of the dijet invariant mass distribution \cite{Bhattiprolu:2020zoq}.
Also, bound states of two $\Theta$'s, produced by gluon exchanges, have a mass slightly below $2 M_\Theta$ and may have phenomenological consequences \cite{Bai:2010dj, Kim:2008bx}.

\begin{figure}[t!]
\hspace*{.4cm}  \includegraphics[width=15cm, angle=0]{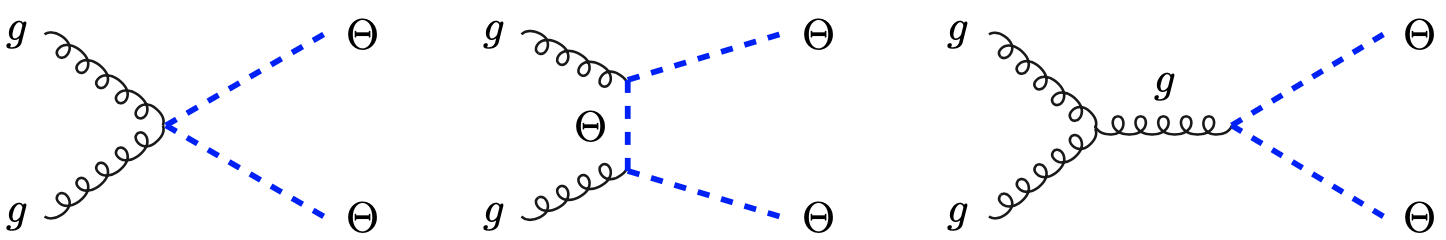}
\vspace*{.2cm}  
\caption{Diagrams responsible for the leading-order $p \, p \to \Theta \, \Theta$ process with initial state gluons. At $\sqrt{s} = 13$ TeV, these account for 91\% of the LO production cross section when  the 
mass of the color-octet scalar $\Theta$ is $M_\Theta = 1$ TeV (the remaining 9\% is due to $q\bar q$ initial states from a diagram similar with the last one shown here).
\vspace*{2mm}
} 
\label{fig:ThetaThetaGluons}
\end{figure}

A larger and more model-independent production mechanism at hadron colliders, is $p p \to \Theta \Theta$ 
via the couplings to gluons. The leading order (LO) pair production process occurs from $gg$ initial state, through the diagrams shown in Figure~\ref{fig:ThetaThetaGluons}, or from $q\bar q$ initial states, through the last diagram of Figure~\ref{fig:ThetaThetaGluons} with the initial gluons replaced by $q$ and $\bar q$. The $p p \to \Theta\Theta$ cross section at LO  ($\sigma_{\rm LO}$) at the LHC is dominated by the $gg$ initial state: 
for $M_\Theta = 1$ TeV,  91.5\% (91.8\%) of $\sigma_{\rm LO}$  
is due to gluons at $\sqrt{s} = 13$ TeV (13.6 TeV), while for $M_\Theta = 2$ TeV the gluon contribution  decreases to 83.6\% (84.3\%).

\begin{figure}[t!]
\hspace*{2.5cm}  \includegraphics[width=11cm, angle=0]{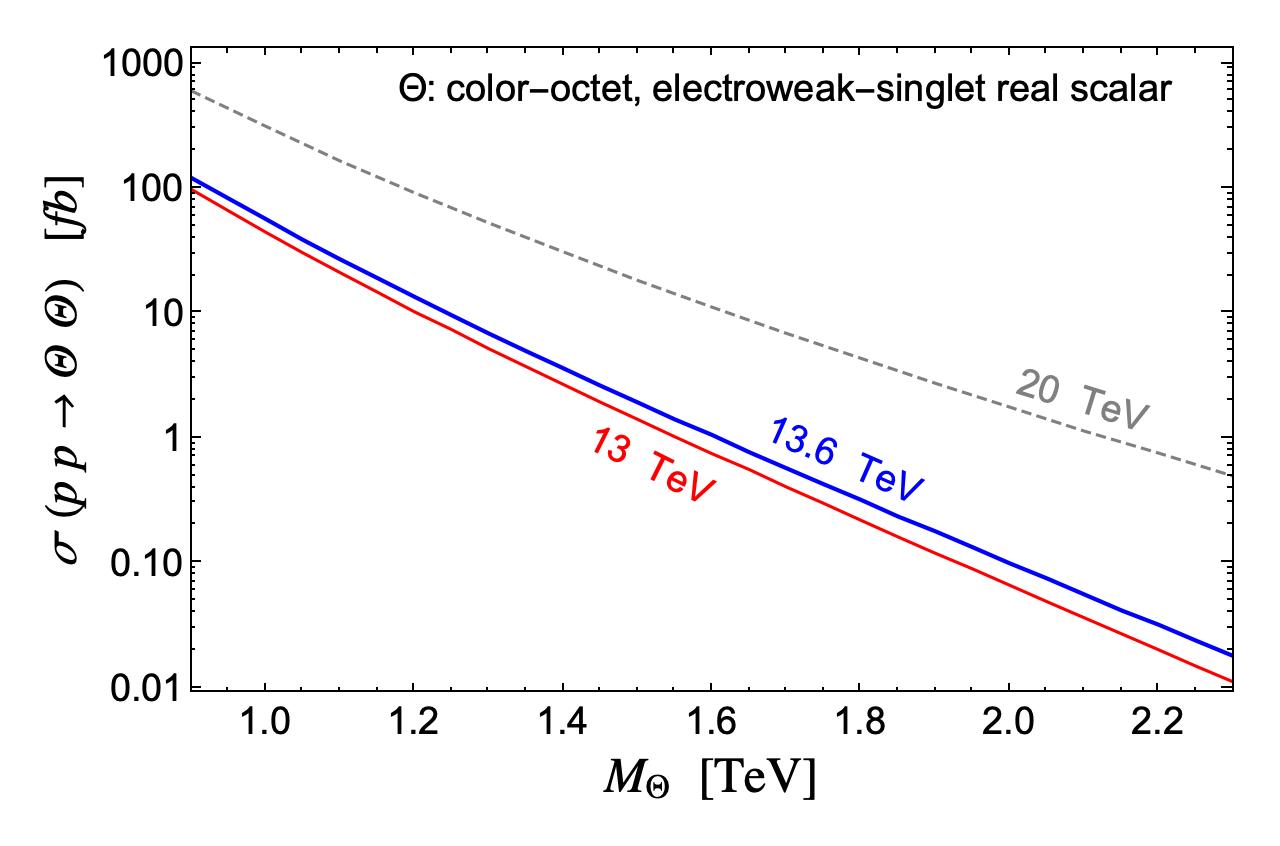}     
\vspace*{-.2cm}  
\caption{Cross section for $p \, p \to \Theta \, \Theta$ computed with Madgraph \cite{Alwall:2014hca} at NLO, at center-of-mass energies of 13 TeV (dashed red line), 13.6 TeV (solid blue line)  and  20 TeV (gray dotted line).
As the process relies on the QCD couplings of the gluons, the cross section depends only on the 
mass of the color-octet real scalar $\Theta$. 
\vspace*{2mm}
} 
\label{fig:xsecTheta}
\end{figure}

Figure~\ref{fig:xsecTheta} shows the cross section for $pp \to \Theta\Theta$ 
computed at NLO with Madgraph \cite{Alwall:2014hca} (which involves various tools, including MadLoop \cite{Hirschi:2011pa}, MadFKS \cite{Frederix:2009yq}, CutTools \cite{Ossola:2007ax}, Ninja \cite{Hirschi:2016mdz}), using model files generated with FeynRules \cite{Alloul:2013bka}  (with loop computations involving NLOCT \cite{Degrande:2014vpa} and FeynArts \cite{Hahn:2000kx}). 
The PDF set used here is the default one employed by Madgraph: 
NNPDF23\_nlo\_as\_0119\_qed  \cite{Ball:2013hta}.
The center of mass energies considered in Figure~\ref{fig:xsecTheta} are 13 TeV (Run 2 of the LHC), 13.6 TeV (Run 3 of the LHC), and 20 TeV (some possible future upgraded LHC or new hadron collider).
Some details about the NLO corrections to $pp \to \Theta\Theta$, including the 1-loop diagrams and the diagrams responsible for 
real emission, are analyzed in  \cite{Goncalves-Netto:2012gvn}, where the case of the octet complex scalar ({\it i.e}., two $\Theta$'s of same mass, or equivalently a sgluon) is considered.

Pair production of $\Theta$ at hadron colliders, followed by each $\Theta$ decaying into two gluons, would lead to a pair of $gg$ resonances mass near $M_\Theta$. A dedicated search of this type has not yet been performed. In Section \ref{sec:2ormore} we will show that the related search for a pair of stops in the final state with four antiquark-initiated jets \cite{CMS:2022usq} leads to different invariant mass distributions, and thus it is not straightforward to translate the limits set there to the four gluon-jet final state.

\medskip

\subsection{Renormalizable origins of $\Theta$ couplings to quarks} 
\label{sec:SMTc} 

Effective couplings of $\Theta$ to the SM quarks ($d^i$ and $u^j$, where $i,j=1,2,3$ are generation indices), of the type
\be
y_{_{\Theta d}}^{ij}  \Theta^a \, \overline d^i_{_L} T^a d_{_R}^j  + y_{_{\Theta u} }^{ij}  \Theta^a \,  \overline u_{_L}^i  T^a u_{_R}^j   +   {\rm H.c.}
\label{eq:eff_Yuk}
\ee
with $T^a$ the generators of the fundamental representation of the color $SU(3)_c$ group, 
are invariant under the $SU(3)_c \times U(1)_{\rm em} $ gauge group, where $U(1)_{\rm em}$ is  the electromagnetic  group. However, the chiral structure of the above couplings, which is enforced by Lorentz symmetry, implies that those Lagrangian terms are not invariant under the electroweak symmetry. 

Thus, to preserve the SM gauge symmetry, the couplings (\ref{eq:eff_Yuk}) must arise from higher-dimensional interactions involving Higgs fields. At dimension 5, only the following operators are possible:
\be
 \frac{ C_{_{\Theta d}}^{ij}  }{ M_* }  \, H^\alpha \, \Theta^a \, \overline q_{_L}^{\alpha i} T^a d_{_R}^j  
 +  \frac{ C_{_{\Theta u}}^{ij}  }{ M_* }  \,  \epsilon^{\alpha\beta} H^{\dagger\alpha} \, \Theta^a \,  \overline q_{_L}^{i \beta} T^a u_{_R}^j  + {\rm H.c.}
\label{eq:eff_ops}
\ee
Here $H$ is the SM Higgs doublet,  $q_L^i = (u_L^i, d_L^i)^\top$ is the quark doublet of generation $i$, $\alpha, \beta = 1,2$ are $SU(2)_W$ indices, and $\epsilon^{\alpha\beta}$ are the $SU(2)_W$ structure functions. The coefficients of the above dimension-5 operators involve complex dimensionless parameters 
$C_{_{\Theta u}}$, $C_{_{\Theta d}}$  ($3\times 3$ matrices in flavor space), and 
are suppressed by the mass $M_* $ of the heavy new particle integrated out to generate those operators.
The parameters of (\ref{eq:eff_Yuk}) and  (\ref{eq:eff_ops}) are related by 
$y_{_{\Theta d}}^{ij} = C_{_{\Theta d}}^{ij} v_{_H} /M_*$ and  $y_{_{\Theta u}}^{ij} = C_{_{\Theta u}}^{ij} v_{_H} /M_*$, where $v_{_H}  \approx 174$ GeV is the electroweak scale.
Operators similar to (\ref{eq:eff_ops}) but involving additional $H^\dagger H$ fields arise at dimension 7 or higher, and we will not discuss them  further.

\begin{table}[t!]
\begin{center}
\renewcommand{\arraystretch}{1.7}
\begin{tabular}{|c|c|c|c|}
\hline 
Field name and label & spin & $\! \! \! $ SM charges $\! \! \! $   & Interactions    
\\
\hline
octo-doublet  \ $\Theta_{\rm _D}$  & 0 & $(8,2, -1/2)$    &  $\!  \left( \mu_{\rm _D}  H  \,  \Theta^a  + y_{_{\rm D}  {_u}} \overline q_{_L}  T^a u_{_R}  + y_{_{\rm D} {_d}}  \overline d_{_R}  T^a q_{_L} \right) \Theta_{\rm _D}^a  \! $   \\
Vquark doublet   \ $\Upsilon$ &  1/2  & $(3,2, +1/6)$  &  $  \!\! \left( y_{u _\Upsilon } H   \overline u_{_R}  \! + y_{d _\Upsilon } H^\dagger  \overline d_{_R} \! \right) \! \Upsilon_{\! _L}  \! \!  + \!  y_{_\Theta {_\Upsilon}  }   \Theta^a  \, \overline q_{_L}  T^a \Upsilon_{\! _R}  \!\!$   \\
up-type  Vquark   \  $\chi$  &  1/2   & $(3,1, +2/3)$  &  $y_{q \chi }   H^\dagger   \, \overline q_{_L}  \chi_{_R} + y_{_\Theta \chi }   \Theta^a  \, \overline u_{_R} T^a \chi_{_L} $   \\
$\! \! \! $ down-type  Vquark \  $\omega$  $\! \! \! $ &  1/2  & $(3,1, -1/3)$   &  $y_{q \omega }   H   \, \overline q_{_L}  \omega_{_R} + y_{_\Theta \omega }   \Theta^a   \, \overline d_{_R}  T^a \omega_{_L} $   \\
\hline
\end{tabular}
\end{center}
\vspace*{-0.3cm}
	\caption{The four types of fields, $\Theta_{\rm _D}$,$\Upsilon$,$\chi$, $\omega$, that can mediate at tree level some (or all) of the dimension-5 operators (\ref{eq:eff_ops}), which couple the $\Theta$ scalar to SM quarks. Their interactions (see last column) with the SM Higgs doublet involve either a mass parameter  $\mu_{\rm _D}$ or a generation-dependent Yukawa coupling, $y_{u _\Upsilon}, y_{d _\Upsilon }, y_{q \chi } , y_{q \omega}$  (generation and $SU(2)_W$ indices are not shown here). The couplings of the vectorlike quarks to $\Theta$ are labelled  $y_{_\Theta {_\Upsilon}  }$,  $y_{_\Theta \chi }$, $y_{_\Theta \omega }$, while the couplings of the SM quarks to $\Theta_{\rm _D}$ are $y_{_{\rm D} {_u}} $, $y_{_{\rm D} {_d}} $.
		 \label{tab:UV}
	 }
\end{table}

\begin{figure}[t!]
\begin{center}
\hspace*{0.3cm}  \includegraphics[width=15cm, angle=0]{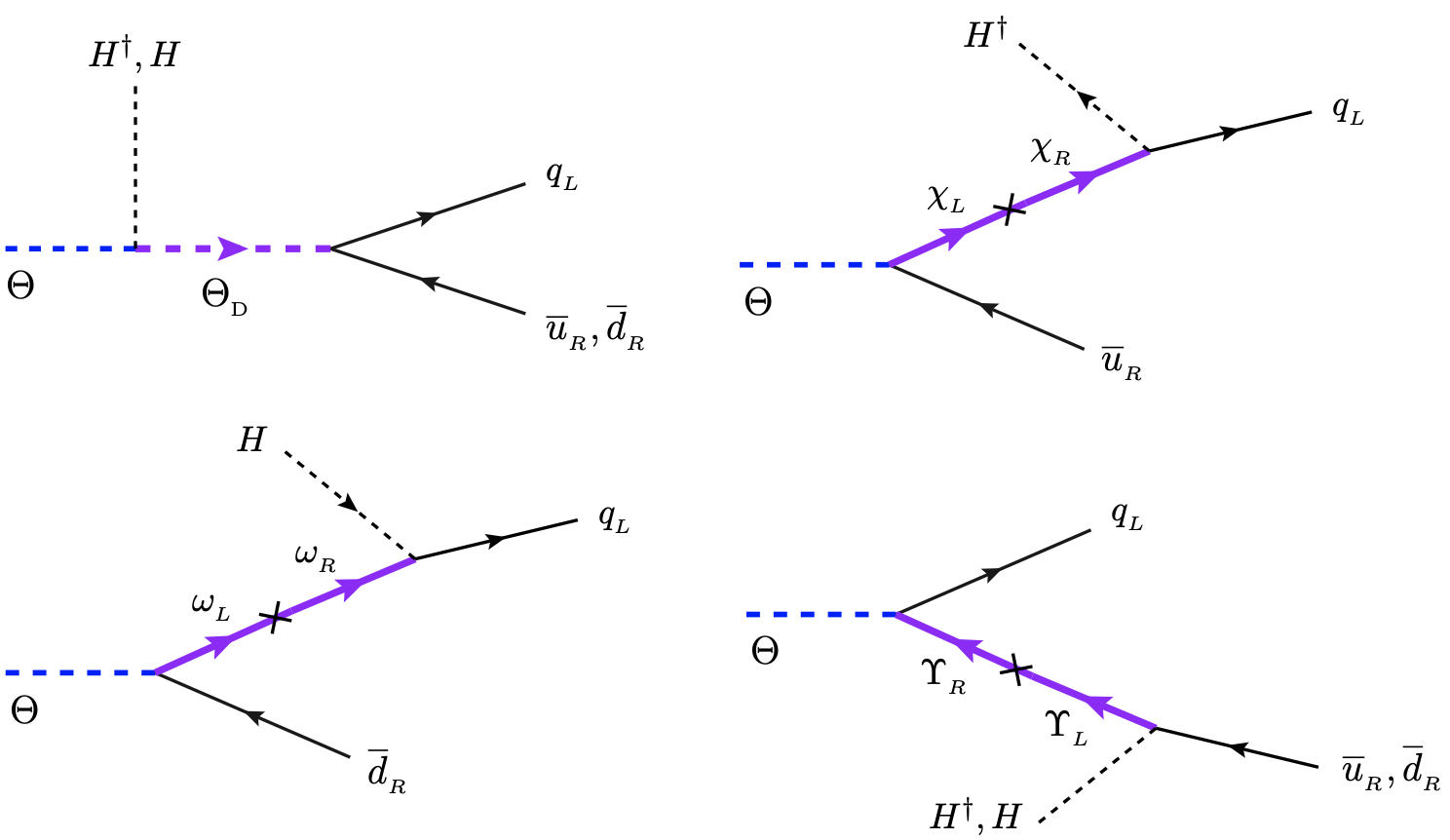}
\end{center}
\vspace*{-0.2cm}  
\caption{Tree-level diagrams that generate the dimension-5 operators (\ref{eq:eff_ops}) when the $\Theta_{\rm _D}$ scalar or any of the $\chi$, $\omega$, $\Upsilon$ Vquarks are integrated out.
\vspace*{5mm}
} 
\label{fig:UVdiagrams}
\end{figure}

Dimension-5 operators (\ref{eq:eff_ops}) may arise from various renormalizable theories, both at tree level and through loops. There are four types of particles, listed with their $SU(3)_c \! \times \! SU(2)_W \! \times \!  U(1)_Y $ gauge charges in Table~\ref{tab:UV}, that can generate the operators  (\ref{eq:eff_ops}) when they are integrated out at tree level. This follows from the four possible attachments of a SM Higgs doublet to a diagram that involves $\Theta$ and a SM quark-antiquark pair, as shown in Figure \ref{fig:UVdiagrams}. 
Thick purple lines in those diagrams represent propagators of the heavy fields that are integrated out. The first diagram is mediated by a color-octet weak-doublet  scalar (referred to here as `octo-doublet'), while the other three are mediated by vectorlike quarks (referred to here as `Vquarks') with different electroweak charges.
These new particles may have masses above the 10 TeV scale, in which case they will not be directly discovered at the LHC, but lower masses (still above $M_\Theta$) are also possible.

Note that the $SU(2)_W$-singlet Vquarks, $\omega$ or $\chi$, mediate only the first or second term in (\ref{eq:eff_ops}), respectively.
In contrast, the $SU(2)_W$-doublet mediators, namely the $\Theta_{\rm _D}$ scalar and the $\Upsilon$ Vquark, may  each mediate all operators in (\ref{eq:eff_ops}).
Thus, four simple renormalizable theories, each involving a single heavy new field beyond $\Theta$ plus the SM,
 generate at tree level at least some of the effective $\Theta$ couplings to the SM quarks  (\ref{eq:eff_Yuk}).

Lagrangian terms containing interactions present in the diagrams of Figure~\ref{fig:UVdiagrams} are given in the last column of Table~\ref{tab:UV}. Each of the Yukawa coupling shown there is either a 3-vector or a $3\times 3$ matrix in flavor space. 
To avoid a lengthy discussion of constraints from flavor-changing processes, we assume that the Yukawa couplings from 
Table~\ref{tab:UV} are negligible in the case of SM quarks of second and third generations.
This assumption may be a consequence of an approximate symmetry, as we show next.

Within the theory that includes just the SM, $\Theta$ and $\Theta_{\rm _D}$, consider for example  
a $Z_4$ symmetry under which $\Theta$ and $\Theta_{\rm _D}$
have charges $+2$, $u_{_R}^1$ has charge $+1$, and $q_{_L}^1$ has charge $+3$,  while the other SM fields are singlets. If the up and down quark masses are generated by the SM Higgs doublet,
then the $H \overline u_{_R}^1 q_{_L}^1$ and $H^\dagger \overline d_{_R}^1 q_{_L}^1$ Yukawa terms explicitly break  the  $Z_4$ symmetry. Nevertheless, the very small $u$ and $d$ masses make this breaking inconsequential: the $\Theta_{\rm _D}$ couplings to second- or third-generation quarks are induced by loops, and are suppressed at least by  $y_{_{\rm D}  {_u}}  m_u v_H/ M_{\Theta_{\rm _D}}^2$ (or the similar expression for the down quark)  where $M_{\Theta_{\rm _D}} \gg  1$ TeV is the octo-doublet mass. Thus, the effective coupling of $\Theta_{\rm _D}$ to the up quark can be several orders of magnitude larger than the ones to any other SM quarks. This is in stark contrast to the color-octet weak-doublet field considered in \cite{Manohar:2006ga,Gresham:2007ri, Gerbush:2007fe, Arnold:2011ra}, which has largest couplings to the third generation quarks based on a minimal-flavor violation hypothesis.

Similar mechanisms are present in the case of the theories that include just the SM, $\Theta$ and one of the Vquarks.
For example, within the theory that includes $\chi$, a $Z_4$ symmetry under which $\Theta$ has charge $+2$, $\chi$ and $q_{_L}^1$ have charge $+1$, and $u_{_R}^1$ has charge $+3$ allows the second diagram of Figure~\ref{fig:UVdiagrams} only for the first generation quarks. As in the above discussion of the octo-doublet, the $Z_4$ symmetry is explicitly broken by the very small Higgs couplings to the SM quarks of the first generation. 
Although we focus here on light quarks in the final state, decays of $\Theta$ to $t \overline t$ or $b \overline b$ are also interesting. The production of a pair of $b \overline b$ resonances was studied in \cite{Bai:2010dj} based on the $\Theta \to b \overline b$ process mediated by an $\omega$ Vquark.

The $Z_4$ symmetries invoked here allow all the Lagrangian terms that involve $\Theta$ in (\ref{eq:lagrangian}) and (\ref{eq:potential}) except for the trilinear $\Theta$ term. As a result, the four types of renormalizable theories discussed here allow $\Theta$ decays to light quarks while preventing the $\Theta$ decay to gluons through a $\Theta$ loop. Thus, the $\Theta \to q \overline q $ decay is by far the dominant one in these theories, and the main LHC signal is $p p \to \Theta\Theta \to (q \bar q)(q \bar q)$. The width-to-mass ratio of $\Theta$ is given in this case by 
\be
\frac{1}{M_\Theta} \, \Gamma (\Theta \to q\bar q) \, \approx \, \frac{y_{_{\Theta q} }^2 }{12 \pi } ~~,
\label{eq:width_qqbar}
\ee
where $y_{_{\Theta q}}$ is the effective Yukawa coupling for the quark $q$ in (\ref{eq:eff_Yuk}). Notice that $y_{_{\Theta q}} \ll 1$, as it arises from integrating out one of the heavy fields from Table~\ref{tab:UV}. 

Although we focus in this paper on final states with light quarks, we emphasize that the couplings of $\Theta$ to the top quarks are free parameters. 
If the $\Theta \to t\bar t$ or $\Theta \to t\bar u$ branching fractions are larger than a few percent, then the processes $p p \to \Theta\Theta \to (t \bar t)(q \bar q)$ or $p p \to \Theta\Theta \to (t \bar u)(q \bar q)$ 
provide interesting signals: a dijet resonance plus a $t \bar t$ or $t \bar u$ resonance of equal mass. 
 If the top quark is sufficiently boosted and decays through a hadronic $W$ channel, 
 then the $(t \bar u)(q \bar q)$ system could be reconstructed as a trijet resonance (if the two jets from the $W$ decay are not resolved) 
 plus a dijet resonance of equal mass.  A similar trijet-dijet final state could arise from $(t \bar t)(q \bar q)$ when one top is highly boosted and the other one is partially resolved. 
Related possible signals (with much smaller branching fractions in the models studied here) include $(t\bar{t})(t\bar{t})$ \cite{Calvet:2012rk,Darme:2018dvz,Beck:2015cga},  $(t \bar c)(t \bar c)$ \cite{Chivukula:2013hga,Calvet:2012rk},  $(t \bar c)(gg)$ \cite{Drueke:2014pla}, and even more  exotic ones \cite{Chen:2014haa}. 
Also interesting are final states involving $b$ quarks, such as $(b \bar b)(b \bar b)$ \cite{Bai:2010dj} or $(b \bar b)(gg)$ \cite{Bai:2011mr}.

There are additional decays modes of $\Theta$ that could be searched for in experiments despite their small branching fractions. 
In particular, 3-body decays involving a SM Higgs boson, a $W$ or a $Z$ (such as $\Theta \to h^0 q \overline q$) are present in each of these theories. 
To see that, recall that the Higgs doublet that shows up in each of the diagrams of Figure~\ref{fig:UVdiagrams} is a collection of a longitudinal $W$, a Higgs boson, and a longitudinal $Z$.  Leptonic decays of any of these heavy SM bosons would provide a good handle for reducing the backgrounds to processes of the type 
$p p \to \Theta\Theta \to (h^0/W/Z + q \bar q) (q \bar q)$. 

As the $h^0/W/Z$ boson is typically boosted, its hadronic decays may not be resolved without a dedicated search, so $\Theta$ may again appear to decay into a 3-jet final state. Pair production of $\Theta$, with one of them 
 decaying to $q \overline q$ and the other to a $h^0/W/Z$  boson plus a $q \overline q$ pair, may appear as a 5-jet final state, with both a dijet mass and a trijet mass near $M_\Theta$. This is again a trijet-dijet signature (as occurred above in the cases of signals with top quarks).  Searches for this type of 5-jet final states have not been yet performed (a search for a pair of trijets was performed by CMS \cite{CMS:2024ldy}).

\medskip

\section{LHC signals of an octet real scalar} 
\label{sec:950GeV}
\setcounter{equation}{0}

The most sensitive search relevant for the electroweak-singlet color-octet scalar, $\Theta$, has been performed by CMS (the nonresonant search for a pair of dijet resonances of approximately equal mass in \cite{CMS:2022usq}) with an integrated luminosity of 138 fb$^{-1}$ in $pp$ collisions at $\sqrt{s}=13$ TeV. 
A related ATLAS search \cite{ATLAS:2017jnp} used a data set that is four times smaller, and thus is less sensitive to possible new physics (the same applies to the earlier CMS search \cite{Sirunyan:2018rlj}). 
The current CMS result shows an excess of 4-jet events that has an average dijet mass $\overline{M}_{jj} \approx 0.95$ TeV, with a local significance of 3.6$\sigma$.
The observed 95\% CL upper limit on the cross section  times 
 the total branching fraction ({\it i.e.}, the product $B_{\rm tot}$ of all branching fractions) 
 times acceptance of the 4-jet final state ($A_{4j}$) is 
  \be 
 \sigma  \, B_{\rm tot} \, A_{4j} < 8.5  \;  \rm fb  ~~,
 \label{eq:CMSlimit}
 \ee
 at $\overline{M}_{jj} = 0.95$ TeV, while the expected limit was approximately $3.3 $ fb.

This reported excess of dijet pairs is primarily observed for $0.94 ~{\rm TeV} < \overline{M}_{jj}<1 ~{\rm TeV}$ in the  $0.25< \overline{M}_{jj}/M_{4j}<0.35$ bin  at a significance of about $3\sigma$, with neighboring $\overline{M}_{jj}$ bins exhibiting smaller excesses of about $1\sigma$.
This excess may be consistent with  a signal from  $\Theta$ pair production. To test this hypothesis, we fix the  color-octet scalar mass in this Section at $M_\Theta = 0.95$ TeV. 
For that mass, the cross section computed with \texttt{Madgraph}~\cite{Alwall:2014hca} at NLO (see Figure~\ref{fig:xsecTheta} and Section~\ref{sec:SMT}  for more details about  the computation)  is  $\sigma (pp \to \Theta\Theta) \approx 65$ fb, with scale uncertainties of about $17\%$.

\begin{figure}[t!]
\hspace*{1cm}  \includegraphics[width=13.1cm, angle=0]{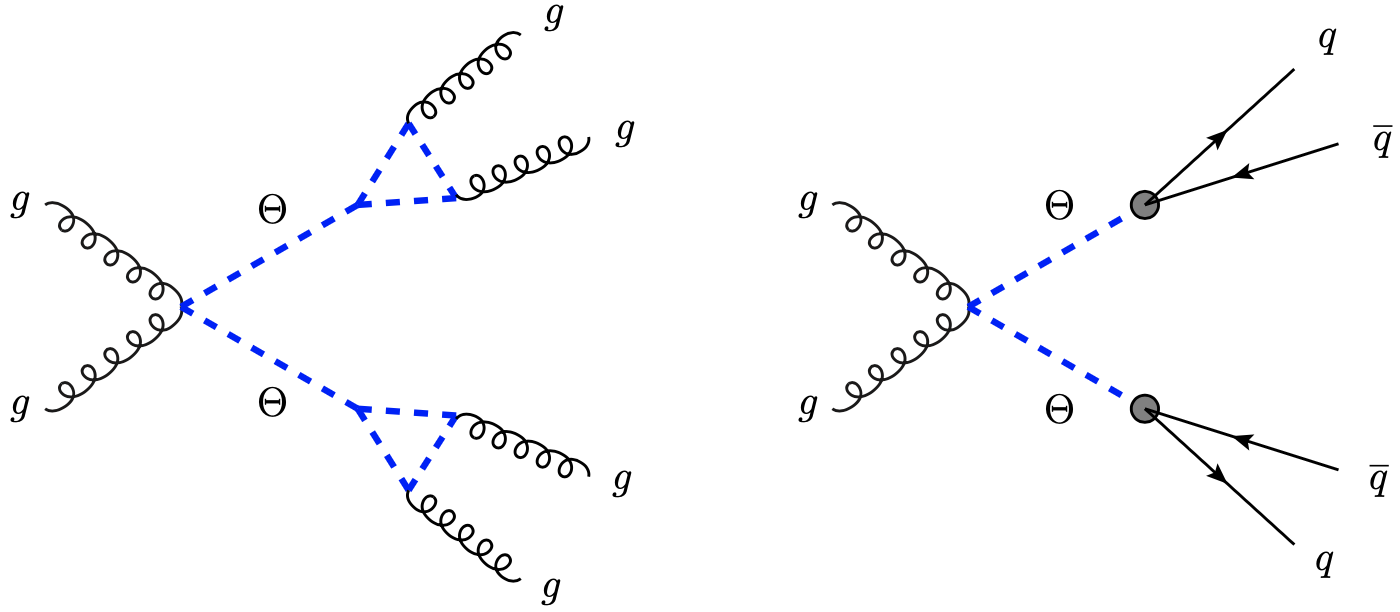}
\vspace*{-0.1cm}  
\caption{Diagrams for the parton-level 4-jet processes due to production of a $\Theta$ pair. The production part of each diagram 
represents only one of the four tree-level processes responsible for $p \, p \to \Theta \, \Theta$ (see Figure~\ref{fig:ThetaThetaGluons}). 
Left diagram involves only the $\Theta$ decay into gluons through a $\Theta$ loop discussed in Section~\ref{sec:SMT}, while 
the right diagram involves only the $\Theta$ decay into quark-antiquark pairs through a higher-dimensional operator (the dark blob contains any of the diagrams in Figure~\ref{fig:UVdiagrams}), as explained in Section~\ref{sec:SMTc}. 
\vspace*{2mm}
} 
\label{fig:diagramTheta4j}
\end{figure}

We need to compute the $A_{4j}$ acceptance for the signal using  the event selection employed by CMS in the nonresonant search \cite{CMS:2022usq}. 
 We consider two possibilities for the decays of the color-octet scalar:  
 either it decays to gluons through a loop (see Section~\ref{sec:SMT} and left diagram of Figure~\ref{fig:diagramTheta4j}) with a branching fraction ${\cal B} (\Theta \to gg) \approx 1$, or it decays to light quarks with a branching fraction ${\cal B} (\Theta \to q\bar q) \approx  1$ through a higher dimensional operator generated as discussed in Section~\ref{sec:SMTc} and shown in right diagram of Figure~\ref{fig:diagramTheta4j}.

While the possibility of comparable branching fractions into gluons and quarks cannot be dismissed, it appears unlikely given that the two widths 
depend on different parameters and may vary over many orders of magnitude, as follows from (\ref{eq:width_gg}) and (\ref{eq:width_qqbar}). 
By contrast, ${\cal B} (\Theta \to gg) \approx 1$ is motivated by the case where there is no additional new physics contributing  to $\Theta \to q\bar q$ at tree level or even at one loop,
while  ${\cal B} (\Theta \to q\bar q) \approx 1$ is motivated by new heavy fields that can mediate tree-level $\Theta$ decays in conjunction with  the assumption that the branching fraction into heavy quarks is suppressed by some small couplings of the new fields. 

Next we present the details of our simulation and follow  the event selection  of the CMS analysis from \cite{CMS:2022usq}. Then, in Section \ref{sec:results}, we discuss our results that include the dijet and 4-jet invariant mass distributions for the  $pp\rightarrow\Theta\Theta\rightarrow (jj) (jj)$ processes, as well as their comparison with the CMS data.  

\smallskip

\subsection{Event generation and overall rate}\label{sec:eventgeneration}
\label{sec:analysis}

Using the \texttt{FeynRules} package~\cite{Alloul:2013bka}, we produced a Universal FeynRules Output (UFO) \cite{Darme:2023jdn} model for the  interactions of the color-octet real scalar $\Theta$ 
with the Lagrangian terms shown 
in (\ref{eq:lagrangian}) and (\ref{eq:eff_Yuk}), supplanted with an effective interaction that accounts for the $\Theta$ loop that generates the coupling of two  gluons to one $\Theta$:  $ M_{\rm eff}^{-1} \, d_{abc} \, \Theta^a \, G^{\mu\nu \, b} \, G_{\mu\nu}^c $, where  $G_{\mu\nu}^b$ is the gluon field strength, and $M_{\rm eff}$ is an effective mass parameter chosen to match 
the value of the  width obtained in (\ref{eq:Thetagg}).
With that  UFO model, we simulated events at LO using the  generator  \texttt{Madgraph5\_aMC@NLO}  version 3.6.3~\cite{Alwall:2014hca} for the process $pp\rightarrow\Theta\Theta\rightarrow (jj) (jj)$, with all $j$'s being first gluon jets, and then quark/antiquark jets. The Madgraph default jet precuts where imposed on each of the four jets: $p_{\rm T} >  20$ GeV and $|\eta| < 5$.
The PDF set used in this simulation is the default one in Madgraph for LO computations: NNPDF23\_lo \cite{Ball:2013hta}.

Then, the parton-level events were showered and hadronized in \texttt{Pythia8.313}~\cite{Sjostrand:2007gs}, and a detector simulation was performed  with \texttt{Delphes} 3~\cite{deFavereau:2013fsa} using the standard CMS card. Jets are clustered according to the anti-$k_{\rm T}$ algorithm \cite{Cacciari:2008gp} with cone size $R=0.4$, using \texttt{FastJet} \cite{Cacciari:2011ma}. The result of these steps is a sample of $10^5$ events, each event including four or more jets that must be paired.  A small set of events, with fewer than four reconstructed jets, is  not included in the analysis.

We use the CMS event selection from Ref.~\cite{CMS:2022usq}. Either a cut on the scalar sum of $p_{\rm T}$'s for all jets in the event is imposed, $H_{\rm T} > 1.05$ TeV, or at least one jet with $p_{\rm T} > 550$ GeV is required. 
Jets (output by Delphes and FastJet in our simulation) are required to have $p_{\rm T} > 80$ GeV and $|\eta| < 2.5$.
For each event, the four jets with the largest $p_{\rm T}$ are paired such that the function 
\be
|\Delta R_1-0.8| + |\Delta R_2 - 0.8|
\label{eq:pairing}
\ee
is minimized, where $\Delta R_{1,2}= \sqrt{(\Delta \phi_{1,2})^2 + (\Delta \eta_{1,2})^2} $ is the jet separation between jet candidates in pair 1 and 2, respectively. Following this pairing, events are selected according to three requirements, each being an upper bound on one of the following kinematic variables: $i)$ the jet separation within each dijet, $\Delta R_{1,2}<2$; \ $ii)$ the pseudorapidity separation between the two dijets, $|\eta_1-\eta_2|<1.1$; \ $iii)$ the asymmetry between the invariant masses of the two dijets, $|M_1- M_2| / (M_1+ M_2) < 0.1$.

For $M_\Theta \approx 0.95$ TeV and $\sqrt{s}=13$ TeV, we found that each of these three requirements independently removes 58\%, 
32\%, and 80\%  of the events, respectively. The combined effect of the event selection described above is that $A_{4j} \approx 6.9\%$ of the events are accepted. 
The acceptance is approximately the same for gluon or quark final states, and it is mostly insensitive to $M_\Theta$ in the $0.9 - 1.1$ TeV range. 

We are now in a position to compare the size of the CMS excess (roughly $5.2$ fb, which is the difference between the observed upper limit  (\ref{eq:CMSlimit}) and the expected limit) with the size of our signal:
\be
\sigma (pp \to \Theta\Theta) \; {\cal B}(\Theta \to jj)^2 \, A_{4j} \approx 4.5 \; {\rm fb}  ~~.
\ee
Note that the central value for the signal rate, which has no adjustable parameter once the dijet resonance mass is fixed, is only about 15\% below the observed excess. 

\medskip

\subsection{Invariant mass distributions} 
\label{sec:results} 

\begin{figure}[b!]
\vspace*{0.3cm}
\includegraphics[width=\textwidth]{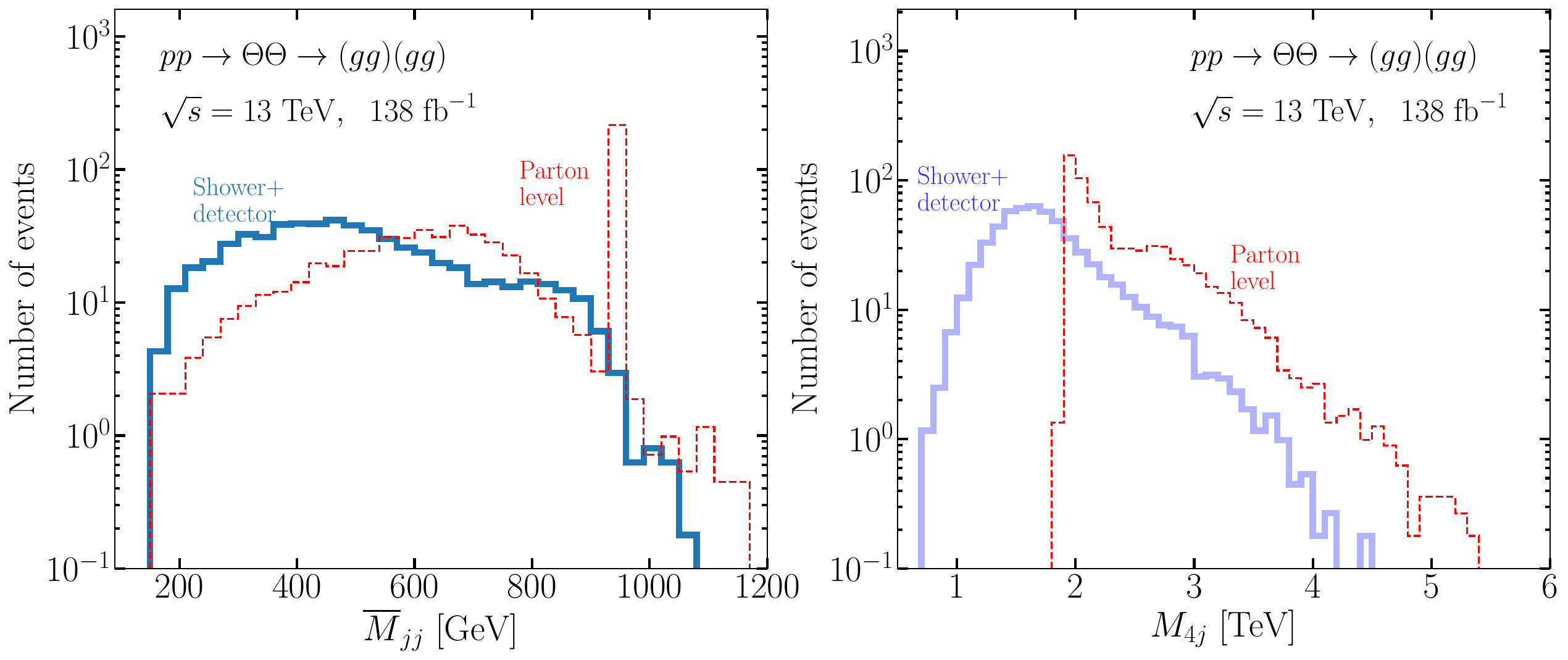}
\vspace*{-7mm}
\caption{Average dijet (left panel) and 4-jet (right panel) invariant mass distributions for $pp\rightarrow \Theta\Theta\rightarrow (gg) (gg)$, for a  color-octet scalar mass $M_\Theta =950$ GeV. Distribution shapes correspond to $\sqrt{s}=13$ TeV, while distribution normalizations are based on 138 fb$^{-1}$ of simulated data and ${\cal B} (\Theta \to gg) = 1$.  Solid blue histogram in each panel shows the signal distribution generated by Madgraph~\cite{Alwall:2014hca} at LO, with showering and detector effects simulated with Pythia~\cite{Sjostrand:2007gs}  and Delphes~\cite{deFavereau:2013fsa}, respectively, using the CMS event selection \cite{CMS:2022usq}. 
For comparison, the dashed red histograms represent the distributions after pairing jets at parton level. The broad left-hand peak in  $\overline{M}_{jj}$ is due to mispaired dijets, and includes mainly events with $M_{4j}$ near 2 TeV.
}
\label{fig:gluondistribution}
\end{figure}

We now turn to more fine-grain details concerning the shapes of the $pp\rightarrow\Theta\Theta\rightarrow (jj) (jj)$ signal distributions for $M_\Theta \approx 0.95$ TeV. 
First,  we use the event sample where all four jets are gluon-initiated. 

In the left-hand (right-hand) panel of Figure~\ref{fig:gluondistribution}, we show the resulting average dijet (4-jet) invariant mass distributions after jet pairing and event selection for an integrated luminosity of 138 fb$^{-1}$ at $\sqrt{s} = 13$ TeV. The solid blue lines represent the distributions after  showering and detector simulation, while the dashed red lines represent the parton-level distributions. The latter are useful for comparison and insight; note that the population of events sharply peaked at an average dijet mass $\overline{M}_{jj}= 0.95$ TeV corresponds to jets that are paired correctly, \textit{i.e.} they each came from the same parent $\Theta$, while the population of events at lower energies corresponds to mismatched jets. The 4-jet invariant mass distribution arising from this nonresonant $\Theta$-pair production exhibits a peak at about $M_{4j}\approx 2M_{\Theta}$, which could be misinterpreted as originating from a broad $s$-channel resonance  (see the recent CMS search  \cite{CMS:2025hpa}). 
The normalization of the distributions in Figure~\ref{fig:gluondistribution} is based on  the central value of the NLO cross section, {\it i.e.,} 65~fb.

\begin{figure}[t!]
\includegraphics[width=\textwidth]{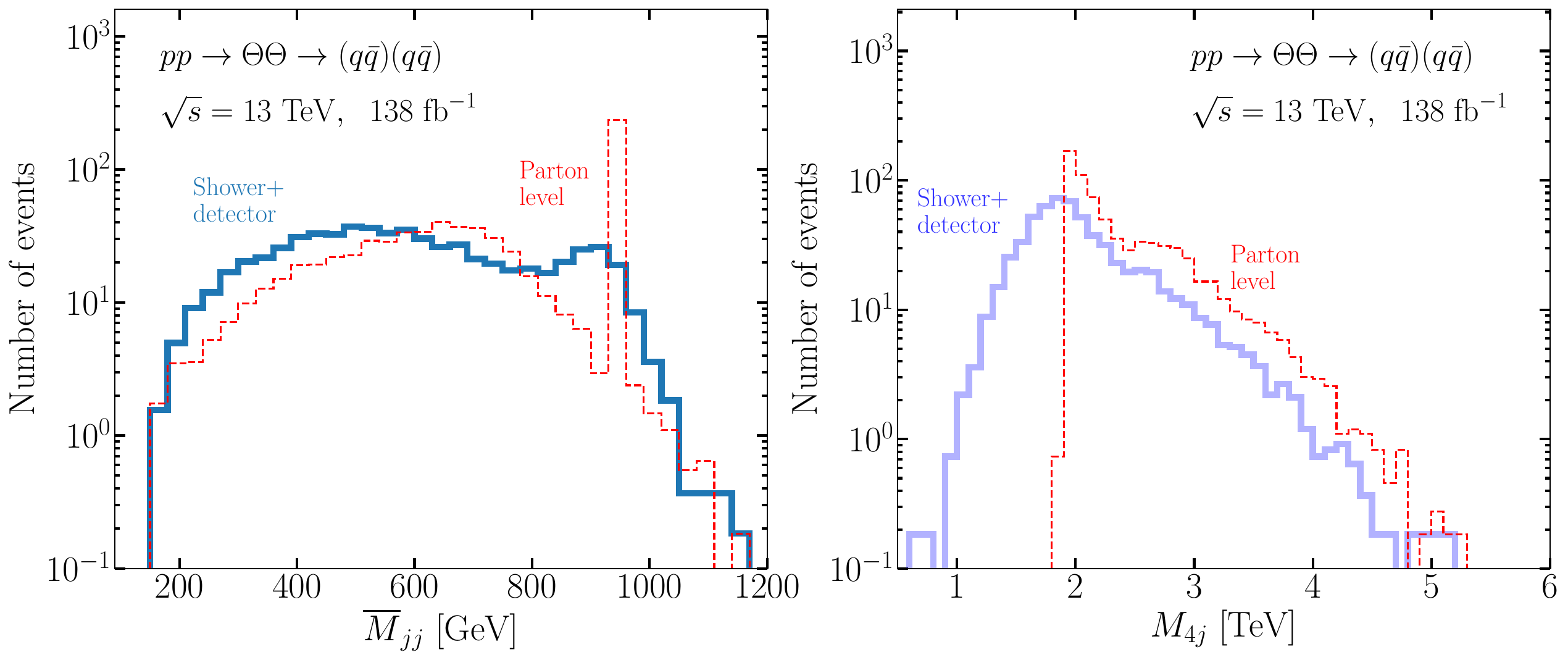}
\vspace*{-7mm}
\caption{Same as in Figure~\ref{fig:gluondistribution}, but for the $pp\rightarrow \Theta\Theta\rightarrow (q \bar q) (q \bar q)$ process with ${\cal B} (\Theta \to q \bar q) = 1$,
where $q$ is a light quark.
Compared to the gluon final state in Figure~\ref{fig:gluondistribution}, these peaks are sharper and narrower near $\overline{M}_{jj} \approx M_\Theta$ and $M_{4j}\approx 2M_\Theta$ due to less radiation emitted outside the jet cone by quarks.
}
\label{fig:quarkdistrib}
\end{figure}

Second, using the event sample where all four jets are initiated by light quarks or antiquarks, we obtain the $\overline{M}_{jj}$ and $M_{4j}$  distributions displayed in Figure~\ref{fig:quarkdistrib}.
Quarks radiate less than gluons, leading to harder jet distributions for quarks  due to more radiation captured in the $R = 0.4$ jet cone. This impacts the shape of the dijet distribution after showering and detector effects (solid blue line in the left panel), in particular near $\overline{M}_{jj} \approx M_{\Theta}$, which has a less suppressed, and narrower peak in comparison to gluon final states. 
A similar effect is seen in the 4-jet distribution for quark final states, shown as a solid blue line in the right panel of Figure~\ref{fig:quarkdistrib}.  
As we will discuss in more detail, the more pronounced peak of the dijet distribution for quark final states makes an important difference when comparing the prediction to the data. For gluons, we find that the peak is too small and broad to explain the excess, and for this reason we will focus on quark jets for the remainder of this Section. 

In the CMS nonresonant analysis~\cite{CMS:2022usq}, events are further divided into three bins of $\overline{M}_{jj}/M_{4j}$:
$(0.15, 0.25)$, $(0.25, 0.35)$ and $(0.35, 0.50)$.
The excess at $\overline{M}_{jj} \approx 0.95$ TeV appears mostly in the second bin, $0.25<\overline{M}_{jj}/M_{4j}<0.35$. 
While the fiducial cross section for $\Theta$ pair production could explain the excess, we further assess the compatibility of the $\Theta$ hypothesis with the data by taking into account the shape of the signal. Thus, we perform a $\Theta$  signal + background fit, with the average dijet invariant mass distribution due to the 4-jet QCD background  modeled (following CMS in \cite{CMS:2022usq}) by the three-parameter  function ``ModDijet-3p", defined by  
\be
\frac{d \, \sigma \!\left(pp \to (jj)(jj) \right) }{d \, \overline{M}_{jj} } = p_0 \, \left(1-x^{1/3} \right)^{p_1} \, x^{-p_2}  ~~,
\label{eq:ModDijet}
\ee
 where $x\equiv \overline{M}_{jj}/\sqrt{s}$, and $p_0, p_1, p_2$ are  parameters fitted to the data ($p_0$ has units of pb/TeV, while $p_1$ and $p_2$ are dimensionless).
  
 Using the CMS $\overline{M}_{jj} $ data points from \cite{HEPDataCMS} (the 
 HEPData~\cite{Maguire:2017ypu} entry for the CMS $4j$ searches \cite{CMS:2022usq}), we perform a $\chi^2$ fit for each bin of $\overline{M}_{jj}/M_{4j}$ separately. We find that the fit of the $\Theta\! \to\! q\bar q$ signal + background hypothesis is better than the background-only hypothesis, with most of the improvement in the $0.25<\overline{M}_{jj}/M_{4j}<0.35$ bin. In the three bins of $\overline{M}_{jj}/M_{4j}$, the values of $\chi^2$ for the background-only hypothesis are $\chi^2_{\rm B} = (25.3, 35.6, 8.4)$,  while for the signal + background fit in the $\Theta \to q\bar q$ case they are smaller: $\chi^2_{\Theta + {\rm B}} =       (24.1, 30.5, 7.7)$. The numbers of degrees of freedom in the three bins are (18, 19, 14).
 
 We note that for gluon jets, the fit does not improve as much over background, with a fit of $\chi^2_{\Theta +B}=(24.9 , 34.6 , 8.1)$. Nevertheless, gluon final states remain important to pursue independently of the current excess near 1 TeV, as they provide a distinct signal shape that might otherwise be overlooked in an analysis. Moreover, the gluon final state is particularly interesting because the  model analyzed in Section~\ref{sec:SMT} involves no free parameters or additional particles beyond $\Theta$.

\begin{figure}[t]
\centering
\includegraphics[width=0.69\textwidth]{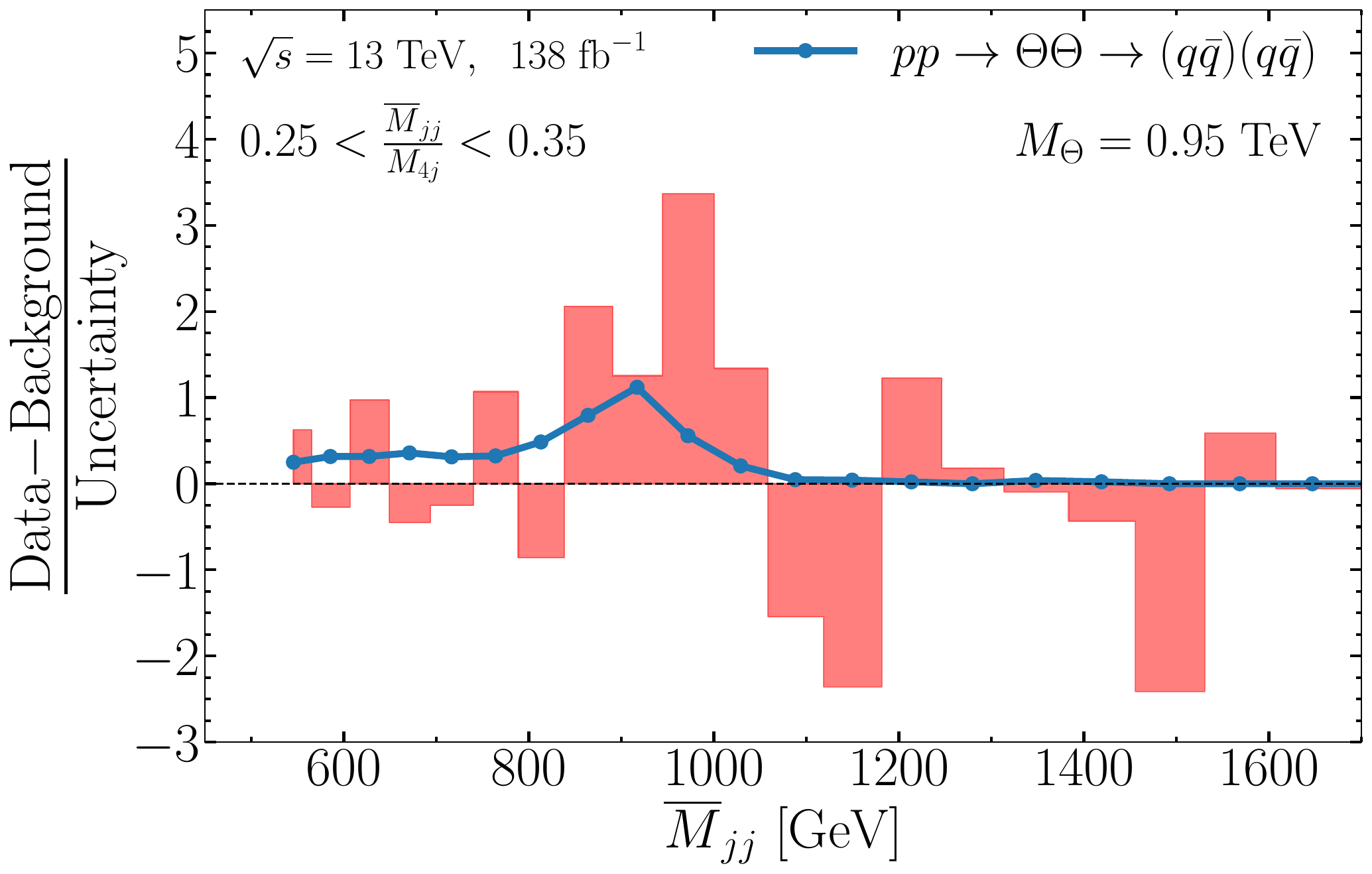}     
\caption{Result of our fit for the average dijet invariant mass distribution ($\overline{M}_{jj}$) arising from the $pp\rightarrow \Theta\Theta\rightarrow (q \bar q) (q \bar q) \to 4j$ signal + background hypothesis,  
for $M_{\Theta}=0.95~{\rm TeV}$ and $0.25<\overline{M}_{jj}/M_{4j}<0.35$.
The red bars show the residuals between the CMS data \cite{CMS:2022usq} and the ModDijet-3p background (\ref{eq:ModDijet})   obtained from our signal + background fit, in units of the CMS experimental uncertainty \cite{HEPDataCMS}; the  $\overline{M}_{jj} \in [0.94 - 1] $ TeV bin exhibits a $\sim \! 3.4\sigma$ excess. The blue curve shows the  signal in the same units. }
\label{fig:residualdistribution}
\end{figure}

In Figure~\ref{fig:residualdistribution} we show the results of our fit. The red bars represent the residuals between the observed data and the background obtained in our fit to the signal + background, divided by the uncertainty in the measurement. Due to the presence of the signal in the modeled distribution, the fitted background is forced to be lower in comparison to the fit from a background-only hypothesis. The result is that the difference between the data and the background  is pushed upwards compared to the background-only residuals shown by CMS in Figure 9 of \cite{CMS:2022usq}. 

If the CMS nonresonant excess is due to $\Theta$, then  the signal should match  the residuals; the solid blue line in Figure~\ref{fig:residualdistribution}  shows our signal divided by the uncertainty \cite{HEPDataCMS}  in the CMS measurement. Given the uncertainties in the data as well as in our simulation and in our fit, the excess may be roughly described by $\Theta$, but a larger signal would be preferable.  
We will explore this possibility in Section~\ref{sec:2ormore}.

The residual plot in Figure~\ref{fig:residualdistribution} is cut off at $\overline M_{jj} \leq 1.7$ TeV. 
There are only two events at larger  $\overline M_{jj}$ in the CMS data \cite{CMS:2022usq}, and both are at $\overline M_{jj} \approx 2$ TeV (in the $0.15<\overline{M}_{jj}/M_{4j}<0.25$ range there is a third event, also near $\overline M_{jj} \approx 2$ TeV). Their origin may itself be some physics beyond the SM: 
a particle of mass around 8 TeV that decays into two particles of mass near 2 TeV \cite{Dobrescu:2018psr, Dobrescu:2019nys, Dobrescu:2024mdl, Bittar:2025rcw}.
We checked that the fit of the background-plus-signal near $\overline M_{jj} \approx 0.95$ TeV does not significantly change  whether or not a new-physics signal is included at $\overline M_{jj} \approx 2$ TeV.

\begin{figure}[t]
\centering
\includegraphics[width=0.6\textwidth]{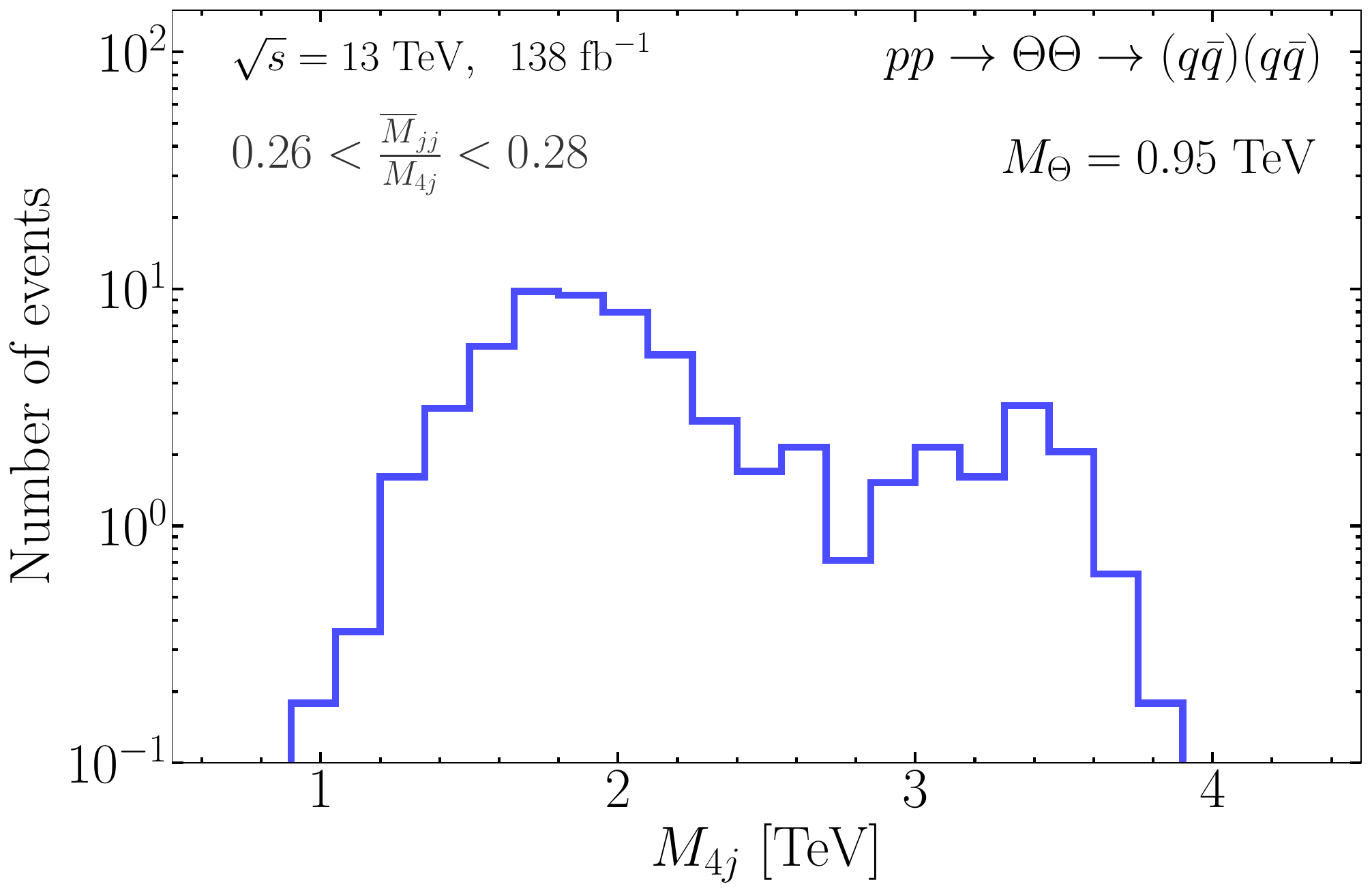}
\caption{Invariant mass distribution for the four quark-jet system in the bin $0.26 < \overline M_{jj}/M_{4j} < 0.28$. 
The mass and collider parameters are chosen here as in Figure~\ref{fig:quarkdistrib}: $M_{\Theta}=0.95~{\rm TeV}$, $\sqrt{s} = 13$ TeV, 138 fb$^{-1}$.
}
\label{fig:4jslice}
\end{figure}

While the excess observed in the average dijet invariant mass distribution provides valuable information for determining its origin, additional information lies in the invariant mass distribution of the four leading jets. In Figure~\ref{fig:4jslice} we show the $M_{4j}$ distribution for $\overline{M}_{jj}/M_{4j} \in [0.26,0.28]$, one of the bins in the broad resonance search performed by CMS~\cite{CMS:2025hpa}. Here, two peak-like structures are visible that may be misinterpreted as coming from two resonances that each decay to four jets. The lower-mass peak at $M_{4j}\approx 2M_\Theta$ is produced near threshold, and is dominated by events that correspond to mispaired jets, while the peak at larger mass, $M_{4j}\approx 3.5 ~{\rm TeV}$, is mostly due to events where $\Theta$ was more boosted, and thus were paired correctly more often. For smaller (larger) values of $\overline{M}_{jj}/M_{4j}$, the lower-mass peak stays in the same position near $M_{4j}\approx 2M_\Theta$, but the higher-mass peak moves towards larger (smaller) values of $M_{4j}$ because it mostly contains events that are correctly paired, so $\overline{M}_{jj}$ is fixed by $M_\Theta$.

It is possible that this information could be leveraged to increase the significance of the dijet excess in case it is due to new physics. If the dijet excess at $\overline{M}_{jj} \approx 0.95~{\rm TeV}$ is due to a new heavy particle, then the kinematic threshold for pair production would be at $M_{4j}\approx 2M_\Theta \approx1.9~{\rm TeV}.$ We find that that majority of signal events with $M_{4j}\lesssim2~{\rm TeV}$ have mispaired dijets; this is because the four jets emerge more isotropically for events produced near threshold and the pairing algorithm based on Eq.~(\ref{eq:pairing}) is only sensitive to jet separations. Therefore, a preliminary signal at $\overline{M}_{jj} \approx 0.95~{\rm TeV}$ could become more significant by considering the other two possible jet pairings for events produced near threshold. Moreover, background events are less likely than signal events to reconstruct near $\overline{M}_{jj} \approx 0.95~{\rm TeV}$ upon subsequent pairings. 

To do this jet re-pairing, the events that qualify as near-threshold events must be defined. The selection should be specific enough (as it should not affect the dijets that are already correctly paired) but also sufficiently generic (as to allow signal enhancement). In our simulation, we find that about $ 28\%$ of all signal events lie in the $ M_{4j} \in \left[ 1770, 2037\right]$ GeV bin used by CMS in the resonant search from Ref.~\cite{CMS:2022usq}, and that 98\% of them are mispaired. This bin is therefore well-suited for secondary pairing. For this subset of events, the choice between the remaining two pairings could be determined by the one that minimizes 
${\cal A}_{1,2} \equiv |M_1 - M_2|/(M_1+M_2)$ and still remove events that do not satisfy the invariant mass asymmetry requirement ${\cal A}_{1,2}<0.1$.

Following this procedure, we find that for $M_\Theta = 0.95~{\rm TeV}$ the number of events with $\overline{M}_{jj}\in \left[0.94,1\right]$ TeV increases by a factor of $2.0$. To limit background events from migrating to this bin, the mass asymmetry requirement on the re-pairing can be tightened at relatively little cost to the signal. For smaller values of the asymmetry, ${\cal A}_{1,2}  <\{0.07,0.05,0.03 \}$, the increase in signal rate for the same $\overline{M}_{jj}$ bin is $\{1.9,1.8, 1.6 \}$. Given this encouraging result, a more detailed study of the background should be pursued in order to determine how much of it  is  affected by this procedure.
 
\bigskip

\section{Color octets  beyond the real scalar} 
\label{sec:2ormore}
\setcounter{equation}{0}

Let us now turn to renormalizable extensions of the SM that include a collection of two or more color-octet real scalars that are nearly degenerate in mass. 
The simplest model of this type includes a complex scalar, $\thc$, that transforms as $(8,1,0)$ under the SM gauge group.
The real and imaginary parts of that complex scalar $\thr$ and $\thi$, transform as octets under the color $SU(3)_c$ gauge group: 
\be
\thc^a = \thr^a + i \thi^a  ~~,
\ee
where $a = 1,...,8$ is the color index. We assume that the Lagrangian terms involving only $\thc^a$ (and gluons in the case of kinetic terms) is invariant under a global $U(1)_{_{\! \rm C}}$ symmetry, with $\thc^a$ carrying charge $+1$,  and its complex conjugate ($\thc^{a \dagger}$) carrying charge $-1$. 
This global symmetry restricts the interactions of $\thr$ and $\thi$ compared to the case where the Lagrangian includes two copies of the real-scalar Lagrangian (\ref{eq:lagrangian}) plus 
all possible mixed terms, such as trilinear terms of the type $d_{abc} \thr^a \thi^b \thi^c$ or  $d_{abc}  \thr^a \thr^b \thi^c$. 
In particular, $U(1)_{_{\! \rm C}}$ forces the masses of  $\thr$ and $\thi$ to be equal. 

In the absence of other fields beyond the SM, the properties of $\thc^a$ are described at the renormalizable level by the following Lagrangian invariant under $U(1)_{_{\! \rm C}}$:
\be
 \left( D^\mu\thc^a \right)^\dagger\! \left(D_\mu\thc^a  \right)  -  \left( {\widetilde{M}_{\thc}}^{\, 2}  \!\! + \! \lambda_H H^\dagger \! H  \right)  \! \thc^{a \dagger} \thc^a
-  \frac{\lambda_{\thc}}{2}   \thc^{a \dagger} \thc^a  \thc^{b \dagger}\thc^b  
- \frac{\lambda_{\thc}^\prime}{4}   \thc^{a \dagger} \thc^b \thc^{a \dagger} \thc^b  
 ~~.
\label{eq:lagrangian-C}
\ee
In isolation, this Lagrangian would imply that $\thc^a$ is a stable particle. We consider, however, interactions of $\thc^a$ with some new particles as described in Section~\ref{sec:SMTc}, which at energy scales smaller than their mass ($M_\star \gg M_{\thc}$) generate an effective coupling to SM quarks of the first generation:
\be
\frac{C_{_{\Theta d}}^{11} }{M_\star} \, H \, \thc^a \; \overline q_{_{L_1}} \! T^a d_{_R} + {\rm H.c.}  ~,
\label{dim5-C}
\ee
where $C_{_{\Theta d}}^{11}  \ll 1$. This effective interaction is $U(1)_{_{\! \rm C}}$ invariant only if the difference between the global charges of $d_{_R}$ and $q_{_{L_1}}$ equals $-1$, assuming that the Higgs doublet is a $U(1)_{_{\! \rm C}}$ singlet. 
In that case, a SM $H \overline q_{_{L_1}} \! d_{_R}$ Yukawa coupling is not allowed by $U(1)_{_{\! \rm C}}$. Nevertheless, the tiny down quark mass may be easily generated by higher-dimensional operators, which in turn arise from integrating out some new heavy fields ({\it e.g.}, a vectorlike quark, or a doublet scalar).  
The operator (\ref{dim5-C}) may be replaced by a similar dimension-5 operator involving the up quark.

Thus, $\thc \to d\bar d$ is the main decay of the color octet, and its width can be large enough for this decay to be prompt. The cross section for 
$p p \to \thc^\dagger \thc$ is twice as large as that for real-scalar pair production shown in Figure~\ref{fig:xsecTheta}. Given that the NLO cross section at $M_{\thc} = 0.95$ TeV is 
$\sigma(p p \to \thc^\dagger \thc) \approx 130$ fb for $\sqrt{s} = 13$ TeV, and the acceptance is $A_{4j} \approx 6.9\%$, the predicted signal rate is  $\sigma A_{4j} \approx 9.0$ fb. 
This seems to exceed the 95\% CL limit on  $\sigma A$ set by the CMS in the nonresonant search \cite{CMS:2022usq}. However, that CMS limit was derived for a stop ($\tilde t \,$) pair production signal of the type $pp \to \tilde t  \, \tilde t^\dagger \to (\bar d \bar s) (ds) $ with an arbitrary normalization fitted to the data. 

It turns out that the stop pair production process differs in some key aspects from our $p p \to \thc^\dagger \thc \to  (d \bar d) (d \bar d) $ signal. 
First, $\tilde t$ decays into two antiquarks, which complicates the hadronization. Typically, two $q\bar q$ pairs from the initial state would provide the quarks needed to hadronize with the two antiquarks from $\tilde t$ decay, as well as the antiquarks that hadronize with the two quarks from  $\tilde t^\dagger$ decay. It is also possible that a $q\bar q$ pair from the vacuum  hadronizes with an antiquark from $\tilde t$ and a quark from $\tilde t^\dagger$, but this is less likely in the case of boosted stops, as there is shorter time for the hadronization to occur when the separation between the $\tilde t$ and $\tilde t^\dagger$ grows fast.

\begin{figure}[t!]
\centering
\includegraphics[width=0.68\textwidth]{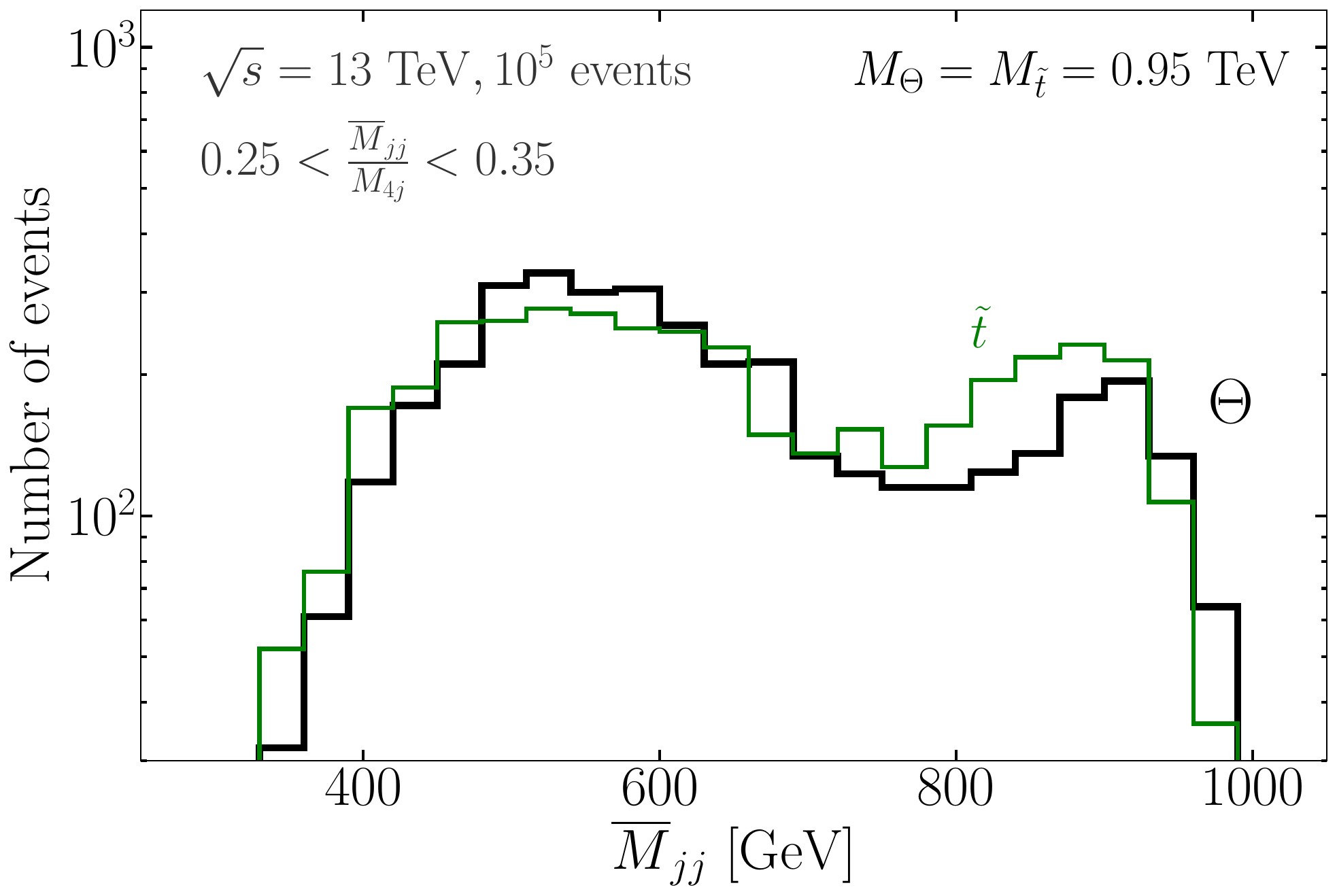}        
\vspace*{-0.3cm}
\caption{Shapes of the  $\overline{M}_{jj}$  distributions for pair production of stops (thin green line) and of color-octet scalars (thick black line), for a mass of 0.95 TeV and $\overline{M}_{jj}/M_{4j}$ within the $(0.25,0.35)$ bin at the 13 TeV LHC. 
}
\vspace*{0.5cm}
\label{fig:stop-shape}
\end{figure}

Therefore, the stop pair production involves long QCD strings between the initial and final states (or between $\tilde t$ and $\tilde t^\dagger$), which means there is more QCD radiation.  As a result, the shape of average dijet invariant mass distribution is flatter and shifted to smaller masses than in the case of color-octet pair production.
This can be seen in Figure~\ref{fig:stop-shape}, where the shapes of the $\overline{M}_{jj}$ distributions generated in the $pp \to \tilde t  \, \tilde t^\dagger \to (\bar d \bar s) (ds) \to 4j$ (thin green line) and $p p \to \thc^\dagger \thc \to  (q \bar q) (q \bar q) \to 4j $ (thick black line) processes are compared. We mention that the $\overline{M}_{jj}$  shape is the same for $\thc^\dagger \thc$ and $\Theta\Theta$ productions, and that the total number of simulated events used there was $10^5$ (before cuts) for the stop-pair process, and the same for  the  $\Theta$-pair process.

We have performed a $\thc$ signal + ModDijet-3p background fit in the $\overline{M}_{jj}$ distribution, and found $\chi^2_{\thc+{\rm B}}=(25.1,27.4,7.8)$ for the three $\overline{M}_{jj}/M_{4j}$ bins (see Section~\ref{sec:results}) used by CMS in the search of nonresonant pairs of dijets  \cite{CMS:2022usq}. 
The fit improves substantially over the background only hypothesis, $\chi^2_{\rm B} = (25.3, 35.6, 8.4)$, especially in the central bin,
$\overline{M}_{jj}/M_{4j} \in (0.25,0.35)$, which contains most of the excess at $\overline{M}_{jj} \approx 0.95$ TeV. 
Note that the fit also improves overall compared to the octet-real-scalar case, $\chi^2_{\Theta + {\rm B}} = (24.1, 30.5, 7.7)$, due to the smaller $\chi^2$ in the central bin, even though the fit is not as good in the side $\overline{M}_{jj}/M_{4j}$  bins.

\begin{figure}[t!]
\centering
\includegraphics[width=0.68\textwidth]{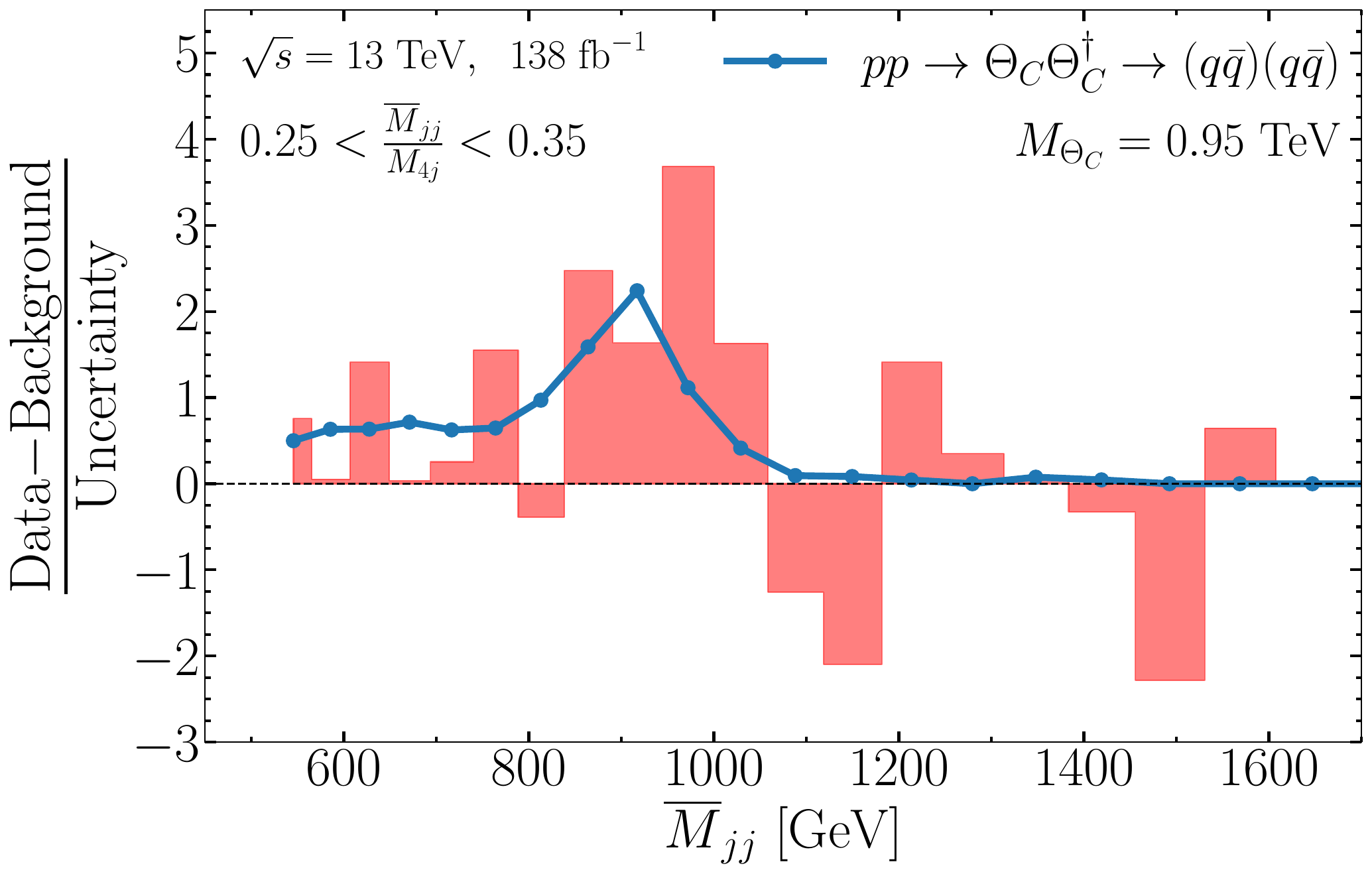}        
\vspace*{-0.3cm}
\caption{Same as Figure~\ref{fig:residualdistribution} except for  $pp\rightarrow \thc\thc^\dagger \rightarrow (q \bar q) (q \bar q) \to 4j$, where the octet $\thc$ is a complex scalar of mass $M_{\thc} =0.95~{\rm TeV}$.
The residuals (red bars) between the CMS data \cite{CMS:2022usq} and the ModDijet-3p background (\ref{eq:ModDijet})   
are shifted up compared to Figure~\ref{fig:residualdistribution} because the signal (solid blue line)  is larger by a factor of 2, so our signal + background fit yielded a smaller background.
}
\vspace*{0.5cm}
\label{fig:residualdistribution2}
\end{figure}

In Figure~\ref{fig:residualdistribution2} we show the result in the $\overline{M}_{jj}$ distribution of the $\thc$  signal + background fit for that $\overline{M}_{jj}/M_{4j}$ bin. Since the signal is twice as large as the one for the octet real scalar, the fitted background is now smaller. Consequently,   
the residuals, shown again as red bars,  between the CMS data \cite{CMS:2022usq} and the background (\ref{eq:ModDijet})  
are shifted up compared to the $\Theta$ signal + background  fit shown in Figure~\ref{fig:residualdistribution}. The largest excess, for
$\overline{M}_{jj} \in [0.94, 1]$ TeV, has now grown to $3.7\sigma$. 

The $p p \to \thc\thc^\dagger \to (q\bar q)  (q\bar q) \to 4j$ signal with $M_{\thc} = 0.95$ TeV is shown as a solid blue line in Figure~\ref{fig:residualdistribution2}. Its peak is at a slightly smaller $\overline{M}_{jj}$ than the bin with the largest observed excess, so one could consider  $M_{\thc}$ values closer to 1 TeV. In that case, it turns out that the fit within the largest $\overline{M}_{jj}/M_{4j}$ bin worsens.
An alternative would be to consider a small breaking of the global $U(1)_{_{\! \rm C}}$ symmetry, giving a mass splitting (from a $\thc\thc+\thc^\dagger\thc^\dagger$ term in the potential)
in the $50-100 $ GeV range between the $\thr$ and $\thi$ components of $\thc$. We will not pursue this possibility here.

The $\Theta$ sector can be even further extended by considering a real scalar that transforms as a triplet under the electroweak symmetry ~\cite{Dobrescu:2011px, CMS:2015fxj} (an `octo-triplet'), labelled here $\tht$. In this case, the Vquark $\Upsilon$ or the octo-doublet $\Theta_{_{\rm D}}$ (see Table~\ref{tab:UV}) can generate the effective interaction that mediates the decay of $\tht$ to quarks. The mass splitting between the physical states of $\tht$ ($\tht^\pm$ and $\tht^0$), induced by loops involving electroweak interactions, is negligible (of order 0.2 GeV~\cite{Dobrescu:2011px}), so the triplet can be seen as a set of three octet real scalars. The cross section is then a factor of 3 larger than that of a single octet real scalar: $\sigma (pp \to \tht\tht^\dagger) \approx 195~{\rm fb}$ and $\sigma (pp \to \tht\tht^\dagger) \, A_{4j}\approx 13.5~{\rm fb}$. The $\tht$ signal has the same shapes in the invariant mass distributions as $\Theta$ or $\Theta_{_{\rm C}}$. 
The fit improves slightly for the central $\overline{M}_{jj}/M_{4j}$ bin, but it worsens for the third bin:  $\chi^2_{\tht+{\rm B}} = (25.1 , 26.5 , 8.5)$.

\section{Conclusions}  \label{sec:conclusions}    \setcounter{equation}{0}

We have investigated the interactions of a color-octet electroweak-singlet scalar, $\Theta$, of mass at the TeV scale. At the LHC, $\Theta$ is pair produced through its gauge coupling to the gluon, such that the total cross section only depends on its mass, $M_\Theta$.  If there are no additional new particles, then $\Theta$ predominantly decays to two gluons at one-loop (see Section \ref{sec:SMT}). The decay width to gluons is highly suppressed, which makes dijet resonance searches blind to single $\Theta$ production through $gg\rightarrow \Theta$. If the model is extended with a certain heavier, decoupled field (a color-octet weak-doublet  scalar or a vectorlike quark, as shown in Table~\ref{tab:UV}), then $\Theta$ decay at tree level to a pair of light quarks may be the dominant decay mode. Thus, the main signature of $\Theta$ at the LHC is a 4-jet final state, where the jets originate from either quarks or gluons.

We consider $\Theta$ in the context of an excess seen in the nonresonant 4-jet analysis from CMS~\cite{CMS:2022usq}. Following an algorithm presented in Section~\ref{sec:950GeV}, the CMS analysis selected pairs of dijets and found a  $3.6\sigma$ excess for an average dijet invariant mass $\overline{M}_{jj}=0.95~{\rm TeV}$. Using color-triplet scalars (stops) as a hypothesis for the shape of the signal (with a floating rate), CMS expected to set a limit on the rate to be 3.3 fb, but the observed limit was set at 8.5 fb. In our work, $\Theta$ pair production was simulated, and the expected rate was found to be 4.5 fb for $M_\Theta=0.95~{\rm TeV}$,  which is close to the size of the excess despite the fact that there are no free parameters. 
Our analysis of the shapes of the $\overline{M}_{jj}$ and $M_{4j}$ distributions shows that 
the distributions predicted from $\Theta$ are roughly in agreement with the shape of the excess, and that quark jets fit the excess better than gluon jets. The jet pairing algorithm mispairs a majority of signal events, and we point out that this may be misinterpreted as a broad 4-jet resonance near $M_{4j}\approx 2M_\Theta$. We also point out modest alterations to the pairing algorithm that would improve the signal significance, but stress that these methods must also be tested on background events. 

In our study of the real scalar $\Theta$, we find that the excess observed in the data suggests that there should be more signal for $\overline{M}_{jj}/M_{4j}\in[0.25, 0.35]$. If the color octet is a complex scalar, or if it is charged under the electroweak symmetry, the additional degrees of freedom result in a larger cross section. 
The rate for a color-octet complex scalar, $\thc$, is $9.0$ fb, which is a bit larger than the observed limit from stop pair production~\cite{CMS:2022usq}. However, due to differences in hadronization between the final states, we find that the $\overline{M}_{jj}$ distribution is distinctly different between stops and octet scalars, such that the limit is not directly applicable. The fit of the $\thc$ signal plus background, shown in Figure~\ref{fig:residualdistribution2}, is in better agreement with the data than for the real scalar.  
We have also considered an octet scalar that is  a weak triplet, and found that as in the case of $\thc$ the fit within the $\overline{M}_{jj}/M_{4j}\in[0.25,0.35]$ bin is   improved, at the expense of a worse fit in the other two bins.

Independently of whether the excess at $\overline{M}_{jj} \approx 0.95~{\rm TeV}$  will be confirmed,  color-octet scalars are compelling candidates for multi-jet searches at the LHC. In particular, they provide a simple origin for a pair of dijet resonances at the TeV scale  in both 4-gluon and $(q\bar q)(q\bar q)$ final states. These invariant mass distributions are different from each other and also different from the ones in the $(qg)(\bar q g)$ final state analyzed in \cite{Dobrescu:2024mdl}.
Moreover, the additional new heavy fields that mediate $\Theta\to q\bar q$ lead to signals that involve a $W$, $Z$ or Higgs boson plus four jets, which in the case of hadronic decays typically give a 5-jet signature of the type trijet-plus-dijet. Related signatures may arise from decays of $\Theta$ involving top quarks.
Pursuing these novel signals could validate the $\Theta$ hypothesis, or may reveal other new phenomena. 
 
 \medskip \bigskip\bigskip


{\bf Acknowledgments:} \ We would like to thank Yang Bai, John Campbell,  Divya Gadkari, Abhijith Gandrakota, Eva Haldakiakis, Robert Harris, Stefan H\"{o}che, Max Knobbe,  Niki Saoulidou, and Ilias Zisopoulos  for insightful comments.
Fermilab is administered by Fermi Forward Discovery Group, LLC under Contract No. 89243024CSC000002 with the U.S. Department of Energy, Office of Science, Office of High Energy Physics.

\smallskip


\end{document}